\documentclass[11pt]{article}
\usepackage{bm}
\usepackage{amssymb}
\usepackage{amsmath}
\usepackage{epsfig}
\usepackage{graphics}
\usepackage{color}
\usepackage{graphicx}
\usepackage{subcaption}
\usepackage{hyperref}
\begin{document}
\begin{center}
{\LARGE\bf MPS: An R package for modelling\\ new families of distributions}\\
\vspace{1cm}
Mahdi Teimouri\\
Department of Statistics\\
Gonbad Kavous University\\
Gonbad Kavous, IRAN\\[2ex]
\end{center}
\vspace{1.5cm}
{\bf Abstract:}~~We introduce an \verb|R| package, called \verb|MPS|, for computing the probability density function, computing the cumulative distribution function, computing the quantile
function, simulating random variables, and estimating the parameters of 24 new shifted families of distributions. By considering an extra shift (location) parameter for each family more flexibility yields. Under some situations, since the maximum likelihood estimators may fail to exist, we adopt the well-known maximum product spacings approach to estimate the parameters of shifted 24 new families of distributions. The performance of the \verb|MPS| package for computing the cdf, pdf, and simulating random samples will be checked by examples. The performance of the maximum product spacings approach is demonstrated by executing \verb|MPS| package for three sets of real data. As it will be shown, for the first set, the maximum likelihood estimators break down but \verb|MPS| package find them. For the second set, adding the location parameter leads to acceptance the model while absence of the location parameter makes the model quite inappropriate. For the third set, presence of the location parameter yields a better fit. 
\section{Introduction}
\setcounter{equation}{0}
Over the last two decades, generalization of the statistical distributions has attracted much attention in the literature. Most of these extensions have been spawned by applications found in analyzing lifetime data. The generalized distributions not only have great potentials to provide families which incorporate more flexible probability density function (pdf), but also exhibit flexible hazard rate function. It is well known that hazard rate function plays the main role in survival and lifetime analysis. Depending on the model which is under study, this function can be constant, decreasing, increasing, upside-down bathtub, and bathtub-shaped. So, the new introduced distributions may have different appeals for different users. 
\par In this work we mainly focus on new generalized families of statistical distributions whose pdf has positive support. Up to now, we are aware of 24 generalized families of distributions with applications in lifetime analysis. We introduce a quite efficient \verb|R| package, called \verb|MPS|, for statistical modelling of 24 generalized families of distributions when they are equipped with the location parameter. The statistical modelling involves computing pdf, computing cumulative distribution function (cdf), simulating random realizations, and estimating the parameters via maximum product spacings (MPS) approach introduced by Cheng and Amin (1983). This paper is organized as follows. In what follows we mention 24 new families of statistical distributions (known in the literature as $G$ families of distributions). A general description about the method of MPS and details for using the \verb|MPS| package for users who are familiar with \verb|R| (\url{R Core Team 2016}) language will be given in Section 2. Section 3 is devoted for checking the \verb|MPS| package through examples and real data applications. We conclude the paper in Section 4.
\par Suppose $G$ is a valid cdf defined on the real line. The general way for introducing a new cdf, $F$ say, is to put the $G$ into the domain of an increasing function such as $h$ with the following form. 
\begin{equation*}
F(x) = h \bigl(G(x,\theta)\bigr),
\end{equation*}
where $h : [0, 1] \rightarrow [0, 1]$ and $\theta$ is parameter space of $G$ distribution. Several candidates exist in the literature for $h$. In the following we review 24 approaches for producing new family of distributions.
\begin{enumerate}
\item Beta exponential $G$ (\verb|betaexpg|) family: Consider the $T-X$ family of distributions introduced by Alzaatreh et al. (2013b). The \verb|betaexpg| family is in fact beta-X family. The cdf and pdf of \verb|betaexpg| family are given as 
\begin{align}\label{betaexpg}
F_{betaexpg}(x,\Theta) &=1 - \frac{{\int_0^{{{\left( {1 - G(x,\theta )} \right)}^d}} {{y^{a - 1}}{{\left( {1 - y} \right)}^{b - 1}}} dy}}{{B(a,b)}},\\
f_{betaexpg}(x,\Theta) &=\frac{{d\,g(x,\theta ){{\left[ {1 - {{\left( {1 - G(x,\theta )} \right)}^d}} \right]}^{b - 1}}{{\left( {1 - G(x,\theta )} \right)}^{ad - 1}}}}{{B(a,b)}},
\end{align}
where $\Theta=(a, b, d, \theta^{T})^{T}$ is the parameter space of \verb|betaexpg| family, $a>0$, $b>0$, and $d>0$ are the new induced shape parameters, and $\theta$ is the parameter space of distribution of $G$. 
\item Beta $G$ (\verb|betag|) family: The \verb|betag| family of distributions introduced by Eugene et al. (2002). The cdf and pdf of the \verb|betag| family are given by
\begin{align}\label{betag}
F_{betag}(x,\Theta) &= \frac{1}{B(a,b)}\int_{0}^{G(x,\theta)}y^{a-1}(1-y)^{b-1}dy,\\
f_{betag}(x,\Theta) &= \frac{1}{B(a,b)}g(x,\theta)\bigl(G(x,\theta)\bigr)^{a-1}\bigl(1-G(x,\theta)\bigr)^{b-1},
\end{align}
where $B(a,b)=\Gamma(a)\Gamma(b)/\Gamma(a+b)$ in which $\Gamma(.)$ is the ordinary gamma function defined as $\Gamma(a)=\int_{0}^{\infty}y^{a-1}e^{-y}dy$ and $\Theta=(a, b, \theta^{T})^{T}$ is the parameter space of the \verb|betag| family. Here, $a>0$ and $b>0$ are the new induced shape parameters and $\theta$ is the parameter space of distribution of $G$. This family can be used for modelling the failure time of a $a$-out-of-$a + b-1$ system when the failure times of the components are independent and identically distributed random variables whose distribution is $G$. Many candidates have been considered in the literature for distribution of $G$ that among them we refer to Pareto \cite{Akinsete2008}, Cauchy \cite{Akinsete2013}, generalized exponential \cite{Barreto2010}, Fr{\'e}chet \cite{Barreto-Souza2011a}, generalized normal \cite{Cintra2014}, Birnbaum-Saunders \cite{Cordeiro2011a}, half Cauchy \cite{Cordeiro2011b}, Laplace\cite{Cordeiro2011c}, power \cite{Cordeiro2012}, moyal \cite{Cordeiro2012b}, extended Weibull \cite{Cordeiro2013c}, generalized gamma \cite{Cordeiro2013a}, generalized Rayleigh \cite{Cordeiro2013b}, exponentiated Weibull \cite{Cordeiro2013c}, Domma \cite{Domma2013}, normal \cite{Eugene2002}, Burr III \cite{Gomes2013a}, inverse Weibull \cite{Hanook2013}, weighted Weibull \cite{Idowu2013}, Gompertz \cite{Jafari2014}, linear failure rate \cite{Jafari2015}, inverse Rayleigh \cite{Leao2013}, Weibull (\cite{Lee2007}, \cite{Famoye2005}), gamma \cite{Kong2007}, Laplace \cite{Kozubowski2008}, generalized Pareto (\cite{Mahmoudi2011}, \cite{Nassar2011}), Lindley \cite{Merovci2014}, lognormal \cite{Montenegro2013}, generalized logistic \cite{Morais2013}, Gumbel \cite{Nadarajah2004a}, Fr{\'e}chet \cite{Nadarajah2004b}, exponential \cite{Nadarajah2006c}, generalized Lindley \cite{Oluyede2015}, Burr XII \cite{Paranaiba2011}, generalized half-normal \cite{Pescim2007}, Lomax \cite{Rajab2013}, Nakagami \cite{Shittu2013}, modified Weibull \cite{Silva2010}, generalized Weibull \cite{Singla2012}, and exponentiated Pareto \cite{Zea2012}.

\item Exponentiated exponential Poisson $G$ (\verb|expexppg|) family: The general form for the cdf and pdf of the \verb|expexppg| family due to Ristic and Nadarajah (2014) are given by
\begin{align}\label{expexppg}
F_{expexppg}(x,\Theta) &= \frac{{1 - {e^{ - b{{\left( {G(x,\theta )} \right)}^a}}}}}{{1 - {e^{ - b}}}},\\
f_{expexppg}(x,\Theta) &= \frac{{a\,b\,g(x,\theta ){{\left( {G(x,\theta )} \right)}^{a - 1}}{e^{ - b{{\left( {G(x,\theta )} \right)}^a}}}}}{{1 - {e^{ - b}}}},
\end{align}
where $\Theta=(a, b, \theta^{T})^{T}$ is the parameter space of the \verb|expexppg| family, $a>0$ and $b>0$ are the new induced shape parameters, and $\theta$ is the parameter space of distribution of $G$. Ristic and Nadarajah (2014) used this family for modelling the time to failure of the first out of a Poisson number of systems functioning independently.

\item Exponentiated $G$ family (\verb|expg|): This family first time introduced by Mudholkar et al. (1991). Contrary to the Weibull distribution that can accommodate just the monotone hazard rates, the hazard rate of the exponentiated Weibull distribution can take unimodal, bathtub shaped, and monotone forms. The general form for cdf and pdf of the \verb|expg| family are given by
\begin{align}\label{expg}
F_{expg}(x,\Theta) &= \bigl(G(x,\theta)\bigr)^{a},\\
f_{expg}(x,\Theta) &= a g(x,\theta)\bigl(G(x,\theta)\bigr)^{a-1},
\end{align}
where $\Theta=(a, \theta^{T})^{T}$ is the parameter space of the \verb|expg| family, $a>0$ is the new induced shape parameter, and $\theta$ is the parameter space of distribution of $G$. This family have been used for several distributions of $G$ among them we refer to Lomax \cite{Abdul-Moniem2012}, modified Weibull \cite{Carrasco2008}, generalized class of distributions \cite{Cordeiro2013d}, generalized Birnbaum-Saunders \cite{Cordeiro2014}, generalized inverse Weibull (\cite{Elbatal2014}, \cite{Gupta1998}, \cite{Gupta1999}, \cite{Lemonte2013b}), Weibull (\cite{Mudholkar1993}, \cite{Mudholkar1995}, \cite{Mudholkar1996}, and \cite{Nadarajah2005}), general exponentiated type \cite{Nadarajah2006a}, Gumbel \cite{Nadarajah2006b}, gamma \cite{Nadarajah2007}, Lomax \cite{Salem2014}, and Pareto \cite{Shawky2009}. 

\item Exponentiated generalized $G$ (\verb|expgg|) family: General form for the cdf and pdf of the \verb|expgg| family are given by
\begin{align}\label{expgg}
F_{expgg}(x,\Theta) &= {\left[ {1 - {{\left( {1 - G(x,\theta )} \right)}^a}} \right]^b},\\
f_{expgg}(x,\Theta) &= a\,b\,g(x,\theta ){\left( {1 - G(x,\theta )} \right)^{a - 1}}{\left[ {1 - {{\left( {1 - G(x,\theta )} \right)}^a}} \right]^{b - 1}},
\end{align}
where $\Theta=(a, b, \theta^{T})^{T}$ is the parameter space of the \verb|expgg| family, $a>0$ and $b>0$ are the new induced shape parameters, and $\theta$ is the parameter space of distribution of $G$. For being familiar with application of this family see \cite{Cordeiro2013d} and \cite{Nadarajah2016}.

\item Exponentiated Kumaraswamy $G$ (\verb|expkumg|) family: Lemonte et al. (2013) introduced \verb|expkumg| family of distributions to model the lifetimes. The cdf and pdf of this family are given by
\begin{align}\label{expkumg}
F_{expkumg}(x,\Theta) &={\left\{ {1 - {{\left[ {1 - {{\left( {G(x,\theta )} \right)}^a}} \right]}^b}} \right\}^d},\\
f_{expkumg}(x,\Theta) &=a\,b\,d\,g(x,\theta ){\left( {G(x,\theta )} \right)^{a - 1}}{\left[ {1 - {{\left( {G(x,\theta )} \right)}^a}} \right]^{b - 1}}{\left\{ {1 - {{\left[ {1 - {{\left( {G(x,\theta )} \right)}^a}} \right]}^b}} \right\}^{d - 1}},
\end{align}
where $\Theta=(a, b, d, \theta^{T})^{T}$ is the parameter space of the \verb|expkumg| family, $a>0$, $b>0$, and $d>0$ are the new induced shape parameters, and $\theta$ is the parameter space of distribution of $G$. Some efforts have been made for investigating the properties and applications of this family. We refer readers to \cite{Huang2014}, \cite{Rodrigues2015}, and \cite{Rodrigues2016}, when distributions of $G$ are supposed to be Dagum, exponential, and inverse Weibull, respectively.

\item Gamma $G$ family (\verb|gammag|): Zografos and Balakrishnan (2009) introduced the \verb|gammag| family of distributions which is similar to that introduced by \cite{Eugene2002} and \cite{Jones2004}. The only difference is that, here, the generator is the cdf of a gamma distribution with shape parameter $a$. General form of the cdf and pdf of \verb|gammag| family are given as
\begin{align}\label{gammag}
F_{gammag}(x,\Theta) &= \frac{\gamma \bigl(-\log(1-G(x,\theta)),a\bigr)}{\Gamma(a)},\\
f_{gammag}(x,\Theta) &= \frac{g(x,\theta)}{\Gamma(a)}{\bigl[-\log(1-G(x,\theta))\bigr]}^{a-1},
\end{align}
where $\gamma (x,a)=\int_{0}^{x}y^{a-1}e^{-y}dy$; for $a>0$, denotes the incomplete gamma function. Now $\Theta=(a, \theta^{T})^{T}$ is the parameter space of the \verb|expkumg| family, $a>0$ is the new induced shape parameter, and $\theta$ is the parameter space of distribution of $G$. The \verb|gammag| family has been studied for several distributions of $G$. Those include Pareto \cite{Alzaatreh2012}, half normal \cite{Alzaatreh2013}, normal \cite{Alzaatreh2014}, exponentiated Weibull \cite{Castellares2015a}, logistic \cite{Castellares2015b}, Dagum \cite{Oluyede2014}, log-logistic \cite{Ramos2013}, and extended Frechet \cite{Silva2013}.

\item Gamma1 $G$ family (\verb|gammag1|): Ristic and Balakrishnan (2012) proposed \verb|gammag1| family of distributions whose cdf and pdf are given by
\begin{align}\label{gammag1}
F_{gammag1}(x,\Theta) &= 1-\frac{\gamma \bigl(-\log(G(x,\theta)),a\bigr)}{\Gamma(a)},\\
f_{gammag1}(x,\Theta) &= \frac{g(x,\theta)}{\Gamma(a)}{\bigl[-\log(G(x,\theta))\bigr]}^{a-1},
\end{align}
where $\Theta=(a, \theta^{T})^{T}$ is the parameter space of \verb|gammag1| family, $a>0$ is the new induced shape parameter, and $\theta$ is the parameter space of distribution of $G$. This family has been studied by \cite{Bhati2015}, \cite{Pinho2012}, and \cite{Pararai2014} and when distributions of $G$ are exponentiated Weibull, inverse Weibull, and Lindley, respectively. 

\item Gamma2 $G$ family (\verb|gammag2|): An extension of \verb|gammag| family, called here \verb|gammag2|, introduced by Torabi and Montazeri (2012). The cdf and pdf of \verb|gammag2| family are given by
\begin{align}\label{gammag2}
F_{gammag2}(x,\Theta) &= \frac{\gamma \Bigl(\frac{G(x,\theta)}{1-G(x,\theta)},a\Bigr)}{\Gamma(a)},\\
g_{gammag2}(x,\Theta) &= \frac{{g(x,\theta )}}{{\Gamma \left( a \right){{\left( {1 - G(x,\theta )} \right)}^2}}}{e^{ - \frac{{G(x,\theta )}}{{1 - G(x,\theta )}}}}{\left( {\frac{{G(x,\theta )}}{{1 - G(x,\theta )}}} \right)^{a - 1}},
\end{align}
where $\Theta=(a, \theta^{T})^{T}$ is the parameter space of \verb|gammag2| family, $a>0$ is the new induced shape parameter, and $\theta$ is the parameter space of distribution of $G$. Torabi and Montazeri (2012) pointed out that the (\ref{gammag2}) family provides great flexibility in modelling the negative and positive skewed, convex-concave shape, and reverse `J' shaped distributions. Also, \verb|gammag2| family has been studied by \cite{Cordeiro2014a} and \cite{Cordeiro2015} when distributions of $G$ are linear failure rate and Lomax, respectively.

\item Generalized beta $G$ family (\verb|gbetag|): Alexander et al. (2012) introduced the \verb|gbetag| family by replacing the generalized beta distribution of the first kind, see \cite{McDonald1984}, with the beta distribution in definition of \verb|betag| family given in (\ref{betag}). The cdf and pdf of \verb|gbetag2| family are given by
\begin{align}\label{gbetag}
F_{gbetag}(x,\Theta) &= \frac{1}{B(a,b)}\int_{0}^{G^d(x,\theta)}y^{a-1}{(1-y)}^{b-1}dy,\\
f_{gbetag}(x,\Theta) &= \frac{d}{B(a,b)}g(x,\theta){\bigl(G(x,\theta)\bigr)}^{ad-1}{\bigl[1-G^d(x,\theta)\bigr]}^{b-1},
\end{align}
where $\Theta=(a, b, d, \theta^{T})^{T}$ is the parameter space of the \verb|gbetag| family, $a>0$, $b>0$, and $d>0$ are the new induced shape parameters, and $\theta$ is the parameter space of distribution of $G$. The \verb|gbetag| family has been studied by for several distributions of $G$. Those include exponential \cite{Barreto2010}, gamma \cite{Marciano2012}, general class of beta generated families \cite{Alexander2012}, Weibull \cite{Singla2012}, Pareto \cite{Mead2014}, and log-logistic \cite{Tahir2014}, and exponentiated gamma \cite{Al-Babtain2015}. 

\item Geometric exponential Poisson $G$ (\verb|gexppg|) family: Nadarajah et al. (2013) introduced the \verb|gexppg| family. The family of distributions proposed by Kus (2007) is special case of \verb|gexppg| family. General form of the cdf and pdf of \verb|gexppg| family are given by
\begin{align}\label{gexppg}
F_{gexppg}(x,\Theta) &= \frac{{{e^{ - a + aG(x,\theta )}} - {e^{ - a}}}}{{{{ {1 - {e^{ - a}} - b + b{e^{ - a + aG(x,\theta )}}} }}}},\\
f_{gexppg}(x,\Theta) &= \frac{{a(1 - b)\,g(x,\theta )\left( {1 - {e^{ - a}}} \right){e^{ - a + a\,G(x,\theta )}}}}{{{{\left( {1 - {e^{ - a}} - b + b{e^{ - a + a\,G(x,\theta )}}} \right)}^2}}},
\end{align}
where $\Theta=(a, b, \theta^{T})^{T}$ is the parameter space of the \verb|gexppg| family, $a>0$ and $0<b<1$ are the new induced shape parameters, and $\theta$ is the parameter space of distribution of $G$. This family is used for modelling the time to failure of the first out of a geometric number of systems functioning independently where the number of parallel units in each system has Poisson distribution and failure times for each units follow independently a $G$ distribution, see \cite{Nadarajah2013a}.

\item Gamma-X family of modified beta exponential $G$ (\verb|gmbetaexpg|) distribution: The cdf and pdf of \verb|gmbetaexpg| family are given by
\begin{align}\label{gmbetaexpg}
F_{gmbetaexpg}(x,\Theta) &={\Bigl( {1 - {e^{ - b\frac{{G(x,\theta )}}{{1 - G(x,\theta )}}}}} \Bigr)^a},\\
f_{gmbetaexpg}(x,\Theta) &=abg(x,\theta ){\left( {1 - G(x,\theta )} \right)^{ - 2}}{e^{ - b\frac{{G(x,\theta )}}{{1 - G(x,\theta )}}}}{\left[ {1 - {e^{ - b\frac{{G(x,\theta )}}{{1 - G(x,\theta )}}}}} \right]^{a - 1}},
\end{align}
where $\Theta=(a, b, \theta^{T})^{T}$ is the parameter space of the \verb|gmbetaexpg| family, $a>0$ and $b>0$ are the new induced shape parameters, and $\theta$ is the parameter space of distribution of $G$. The \verb|gmbetaexpg| is in fact the gamma-X family due to Alzaatreh et al. (2013b). We address readers to Alzaatreh et al. (2012), Alzaatreh and Knight (2013), and Alzaatreh et al. (2014) for properties and applications of the gamma-Pareto, gamma-half normal, and gamma-normal families, respectively.

\item Generalized transmuted-$G$ (\verb|gtransg|) family: The functional combination of the cdf of a given distribution with the inverse cdf of another distribution known in the literature as the transmutation map. The cdf and pdf of the generalized transmuted-$G$, called here \verb|gtransg|, due to Merovci et al. (2017) are given by
\begin{align}\label{gtransg}
F_{gtransg}(x,\Theta) =& {\left( {G(x,\theta )} \right)^a}{\left[ {1 + b\left( {1 - G(x,\theta )} \right)} \right]^a},\\
f_{gtransg}(x,\Theta) =&a\,g(x,\theta ){\left( {G(x,\theta )} \right)^{a - 1}}\left[ {1 + b - 2bG(x,\theta )} \right]{\left[ {1 + b\left( {1 - G(x,\theta )} \right)} \right]^{a - 1}},
\end{align}
where $\Theta=(a, b, \theta^{T})^{T}$ is the parameter space of the \verb|gtransg| family, $a>0$ and $-1<b<1$ are is the new induced shape parameters, and $\theta$ is the parameter space of distribution of $G$. The transmutation map has been applied in the literature to many candidates for distribution of $G$. Those include inverse Rayleigh \cite{Ahmad2014}, extreme value \cite{Aryal2009}, Weibull \cite{Aryal2011}, log-logistic \cite{Aryal2013}, Rayleigh \cite{Merovci2013b}, Lindley \cite{Merovci2013a}, generalized Rayleigh \cite{Merovci2014a}, Pareto \cite{Merovci2014b}, and Lindley-geometric \cite{Merovci2014c}. It should be noted that there are other generalizations of the transmuted-$G$ family including: exponentiated transmuted Weibull \cite{Hady2014}, beta transmuted Weibull \cite{Pal2014}, transmuted exponentiated generalized-$G$ family \cite{Yousof2015}, generalizations of the transmuted-$G$ family \cite{Nofal2016}, transmuted geometric-$G$ family \cite{Afify2016a}, Kumaraswamy transmuted-$G$ family\cite{Afify2016b}, beta transmuted-$G$ family \cite{Afify2017}, and complementary generalized transmuted Poisson-$G$ \cite{Alizadeh2018}. 

\item Log-logistic-X family of $G$ (\verb|gxlogisticg|) distribution: The \verb|gxlogisticg| family is a special case of $T-X$ family due to Alzaatreh et al. (2013b). If we let $T$ follow a log-logistic distribution with shape parameter $a$ and $W(F(.))=-\log(1-F(.))$, then the \verb|gxlogisticg| family is obtained. The cdf and pdf of the \verb|gxlogisticg| family are given by
\begin{align}\label{gxlogisticg}
F_{gxlogisticg}(x,\Theta) &=\frac{1}{{1 + {{\left[ { - \log \left( {1 - G(x,\theta )} \right)} \right]}^{-a}}}}, \\
f_{gxlogisticg}(x,\Theta) &= \frac{{ag(x,\theta ){{\left[ { - \log \left( {1 - G(x,\theta )} \right)} \right]}^{ - a - 1}}}}{{\left( {1 - G(x,\theta )} \right){{\left\{ {1 + {{\left[ { - \log \left( {1 - G(x,\theta )} \right)} \right]}^a}} \right\}}^2}}},
\end{align}
where $\Theta=(a, \theta^{T})^{T}$ is the parameter space of the \verb|gxlogisticg| family, $a>0$ is the new induced shape parameter, and $\theta$ is the parameter space of distribution of $G$.

\item Kumaraswamy $G$ family (\verb|kumg|): Based on Kumaraswamy (1980) distribution, Jones (2009) introduced a new family of distributions which is known as Kumaraswamy $G$ family in the literature. The cdf and pdf of \verb|kumg| family are given by
\begin{align}\label{kumg}
F_{kumg}(x,\Theta) &=1 - {\left[ {1 - {{\left( {G(x,\theta )} \right)}^a}} \right]^b},\\
f_{kumg}(x,\Theta) &= a\,b\,g(x,\theta ){\left( {G(x,\theta )} \right)^{a - 1}}{\left[ {1 - {{\left( {G(x,\theta )} \right)}^a}} \right]^{b - 1}},
\end{align}
where $\Theta=(a, b, \theta^{T})^{T}$ is the parameter space of the \verb|gtransg| family, $a>0$ and $-1<b<1$ are is the new induced shape parameters, and $\theta$ is the parameter space of distribution of $G$. Many candidates have been considered in the literature for distribution of $G$. Among them we refer to modified inverse Weibull \cite{Aryal2015}, Pareto \cite{Bourguignon2013}, Lindley \cite{Cakmakyapan2014}, Weibull (\cite{Cordeiro2010}, \cite{Cordeiro2011}), Gumbel (\cite{Cordeiro2011}, \cite{Cordeiro2012a}), normal \cite{Cordeiro2011}, inverse-Gaussian \cite{Cordeiro2011}, generalized half-normal \cite{Cordeiro2012d}, modified Weibull \cite{Cordeiro2014b}, generalized (Stacy) gamma (\cite{Cordeiro2011}, \cite{de Pascoa2011}), log-logistic \cite{de Santana2012}, generalized linear failure rate \cite{Elbatal2013a}, exponentiated Pareto \cite{Elbatal2013b}, quasi Lindley \cite{Elbatal2013c}, Kumaraswamy \cite{El-Sherpieny2014}, generalized Rayleigh \cite{Gomes2014b}, half-Cauchy \cite{Ghosh2014}, generalized Pareto \cite{Nadarajah2013b}, inverse exponential \cite{Oguntunde2014}, Burr XII distribution \cite{Paranaiba2013}, generalized gamma distribution \cite{Pascoa2011}, inverse Rayleigh \cite{Roges2014}, log-logistic \cite{Santana2012}, Birnbaum-Saunders \cite{Saulo2012}, inverse Weibull \cite{Shahbaz2012}, generalized exponentiated Pareto \cite{Shams2013a}, and generalized Lomax \cite{Shams2013b}.

\item Log-gamma1 $G$ (\verb|loggammag1|) family: This family introduced by Amini et al. (2013). The cdf and pdf of the \verb|loggammag1| family are given by
\begin{align}\label{loggammag1}
F_{loggamma1}(x,\Theta) &=\int_0^{ - b\log \left( {1 - G(x,\theta )} \right)} {\frac{{{y^{a - 1}}{e^{ - y}}}}{{\Gamma (a)}}dy}, \\
f_{loggamma1}(x,\Theta) &= \frac{{{b^a}}}{{\Gamma (a)}}g(x,\theta ){\left[ { - \log (1 - G(x,\theta ))} \right]^{a - 1}}{\left( {1 - G(x,\theta )} \right)^{b - 1}},
\end{align}
where $\Theta=(a, b, \theta^{T})^{T}$ is the parameter space of the \verb|loggammag1| family, $a>0$ and $b>0$ are the new induced shape parameters, and $\theta$ is the parameter space of distribution of $G$. Amini et al. (2013) applied this family to model the earnings of workers from the US Bureau of Labor Statistics.
\item Log gamma type II $G$ (\verb|loggammag2|) family: General form for the cdf and pdf of the \verb|loggammag2| family due to Amini et al. (2013) are given by
\begin{align}\label{loggammag2}
F_{loggammag2}(x,\Theta) &=1 - \int_0^{ - b\log \left( {G(x,\theta )} \right)} {\frac{{{y^{a - 1}}{e^{ - y}}}}{{\Gamma (a)}}dy},\\
f_{loggammag2}(x,\Theta) &=\frac{{{b^a}}}{{\Gamma (a)}}g(x,\theta ){\left[ { - \log (G(x,\theta ))} \right]^{a - 1}}{\left( {G(x,\theta )} \right)^{b - 1}},
\end{align}
where $\Theta=(a, b, \theta^{T})^{T}$ is the parameter space of the \verb|loggammag2| family, $a>0$ and $b>0$ are the new induced shape parameters, and $\theta$ is the parameter space of distribution of $G$. Amini et al. (2013) applied this family to model the earnings of workers from the US Bureau of Labor Statistics.

\item Modified beta $G$ (\verb|mbetag|) family: General form for the cdf and pdf of the \verb|mbetag| family are given by
\begin{align}\label{mbetag}
F_{mbetag}(x,\Theta) &= \frac{{\int_0^{\frac{{d\,G(x,\theta)}}{{1 - (1 - d)G(x,\theta )}}} {{y^{a - 1}}{{\left( {1 - y} \right)}^{b - 1}}} dy}}{{B(a,b)}},\\
f_{mbetag}(x,\Theta) &=\frac{{{d^a}g(x,\theta ){{\left( {G(x,\theta )} \right)}^{a - 1}}{{\left( {1 - G(x,\theta )} \right)}^{b - 1}}}}{{B(a,b){{\left[ {1 - \left( {1 - d} \right)G(x,\theta )} \right]}^{a + b}}}},
\end{align}
where $\Theta=(a, b, d, \theta^{T})^{T}$ is the parameter space of the \verb|mbetag| family, $a>0$, $b>0$, and $d>0$ are the new induced shape parameters, and $\theta$ is the parameter space of distribution of $G$. The \verb|mbetag| family was used to model S\&P/IFC (Standard \& Poor's/International Finance Corporation) global daily price indices in United States dollars for South Africa, see \cite{Nadarajah2014b}. Also, a slightly different of this family is the \verb|betag|-geometric family that has been investigated when distributions of $G$ are exponential (\cite{Adamidis1998}, \cite{Bidram2012}), Kumaraswamy \cite{Akinsete2014}, and Weibull (\cite{Cordeiro2013e}, \cite{Bidram2013}).

\item Marshal-Olkin $G$ family (\verb|mog|): Marshall and Olkin (1997) proposed a new approach for adding a parameter to a family of distributions and then applied it exponential and Weibull families. General form for the cdf and pdf of the \verb|mog| family are given by
\begin{align}\label{mog}
F_{mog}(x,\Theta) &= 1 - \frac{{a\left( {1 - G(x,\theta )} \right)}}{{\left[ {1 - \left( {1 - a} \right)\left( {1 - G(x,\theta )} \right)} \right]}},\\
f_{mog}(x,\Theta) &= \frac{{ag(x,\theta )}}{{{{\left[ {1 - \left( {1 - a} \right)\left( {1 - G(x,\theta )} \right)} \right]}^2}}},
\end{align}
where $\Theta=(a, \theta^{T})^{T}$ is the parameter space of the \verb|mog| family, $a>0$ is the new induced shape parameter, and $\theta$ is the parameter space of distribution of $G$. Rubio and Mark (2012) studied the Marshall and Olkin's (1997) approach as a skewing mechanism. Also, properties and applications of this family have been studied for many distributions of $G$ including extended Burr type XII \cite{Al-Saiari2014}, generalized (Stacy) gamma (\cite{Barrigaa2018}, \cite{de Pascoa2011}), log-logistic \cite{de Santana2012}, exponential Pareto \cite{El-Said2017}, Esscher transformed Laplace \cite{George2013}, extended Weibull \cite{Ghitany2005}, extended Lomax \cite{Ghitany2007}, power log-normal \cite{Gui2013a}, extended log-logistic \cite{Gui2013b}, beta (\cite{Jose2009}, \cite{Nadarajah2014b}), $q$-Weibull \cite{Jose2010}, extended uniform \cite{Jose2011}, Morgenstern Weibull \cite{Jose2013}, generalized asymmetric Laplace \cite{Krishna2011}, Fr{\'e}chet \cite{Krishna2013}, Birnbaum-Saunders \cite{Lemonte2013a}, inverse Weibull \cite{Okash2017}, Zipf \cite{Perez-Casany2014}, gamma \cite{Ristic2007}, and discrete uniform \cite{Sandhya2014}.

\item Marshall-Olkin Kumaraswamy $G$ (\verb|mokumg|) family: General form for the cdf and pdf of the \verb|mokumg| family due to Roshini and Thobias (2017) are given by
\begin{align}\label{mokumg}
F_{mokumg}(x,\Theta) &= 1 - \frac{{d{{\left[ {1 - {{\left( {G(x,\theta )} \right)}^a}} \right]}^b}}}{{1 - \left( {1 - d} \right){{\left[ {1 - {{\left( {G(x,\theta )} \right)}^a}} \right]}^b}}},\\
f_{mokumg}(x,\Theta) &= \frac{{abd g(x,\theta ){{\left( {G(x,\theta )} \right)}^{a - 1}}{{\left[ {1 - {{\left( {G(x,\theta )} \right)}^a}} \right]}^{b - 1}}}}{{{{\left[ {1 - \left( {1 - d} \right){{\left[ {1 - {{\left( {G(x,\theta )} \right)}^a}} \right]}^b}} \right]}^2}}},
\end{align}
where $\Theta=(a, b, d, \theta^{T})^{T}$ is the parameter space of the \verb|mokumg| family, $a>0$, $b>0$, and $d>0$ are the new induced shape parameters, and $\theta$ is the parameter space of distribution of $G$. 
\item Odd log-logistic $G$ (\verb|ologlogg|) family: Gauss et al. (2017) introduced the \verb|ologlogg| family. General form for the cdf and pdf of this family are given by
\begin{align}\label{ologlogg}
F_{ologlogg}(x,\Theta) &=\frac{{a\,b\,d\,g(x,\theta ){{\left( {G(x,\theta )} \right)}^{a\,d - 1}}{{\left[ {\bar G(x,\theta )} \right]}^{d - 1}}}}{{{{\left[ {{{\left( {G(x,\theta )} \right)}^d} - {{\left( {\bar G(x,\theta)} \right)}^d}} \right]}^{a + 1}}}}{\left\{ {1 - {{\left[ {\frac{{{{\left( {G(x,\theta )} \right)}^d}}}{{{{\left( {G(x,\theta )} \right)}^d} - {{\left( {\bar G(x,\theta )} \right)}^d}}}} \right]}^a}} \right\}^{b - 1}},\\
f_{ologlogg}(x,\Theta) &= 1 - {\left\{ {1 - {{\left[ {\frac{{{{\left( {G(x,\theta )} \right)}^d}}}{{{{\left( {G(x,\theta )} \right)}^d} - {{\left( {\bar G(x,\theta )} \right)}^d}}}} \right]}^a}} \right\}^b},
\end{align}
where $\bar G(x,\theta ) = 1 - G(x,\theta )$, $\Theta=(a, b, d, \theta^{T})^{T}$ is the parameter space of the \verb|ologlogg| family, $a>0$, $b>0$, and $d>0$ are the new induced shape parameters, and $\theta$ is the parameter space of distribution of $G$. 

\item Truncated-exponential skew-symmetric $G$ (\verb|texpsg|) family: General form for the cdf and pdf of the \verb|texpsg| family are given by
\begin{align}\label{texpsg}
F_{texpsg}(x,\Theta) &=\frac{{1 - {e^{ - aG(x,\theta )}}}}{{1 - {e^{ - a}}}},\\
f_{texpsg}(x,\Theta) &= \frac{a}{{1 - {e^{ - a}}}}g(x,\theta ){e^{ - aG(x,\theta )}},
\end{align}
where $\Theta=(a, \theta^{T})^{T}$ is the parameter space of the \verb|texpsg| family, $a>0$ is the new induced shape parameter, and $\theta$ is the parameter space of distribution of $G$. This family was used for modelling the annual maximum daily rainfall of 14 locations in west central Florida, see \cite{Nadarajah2014a}. 

\item Weibull extended $G$ (\verb|weibullextg|) family: The \verb|weibullextg| is in fact the Weibull-$X$ family introduced by Alzaatreh et al. (2013b). General form for the cdf and pdf of the \verb|weibullextg| family are given by
\begin{align}\label{weibullextg}
F_{weibullextg}(x,\Theta) &=1 - \exp \left\{ { - a{{\left( {\frac{{G(x,\theta )}}{{1 - G(x,\theta )}}} \right)}^{\frac{1}{b}}}} \right\},\\
f_{weibullextg}(x,\Theta) &=\frac{{a\,g(x,\theta)}}{{b{{\left( {1 - G(x,\theta )} \right)}^2}}}{\left( {\frac{{G(x,\theta )}}{{1 - G(x,\theta )}}} \right)^{\frac{1}{b} - 1}}\exp \left\{ { - a{{\left( {\frac{{G(x,\theta )}}{{1 - G(x,\theta )}}} \right)}^{\frac{1}{b}}}} \right\},
\end{align}
where $\Theta=(a, b, d, \theta^{T})^{T}$ is the parameter space of the \verb|weibullextg| family, $a>0$, $b>0$, and $d>0$ are the new induced shape parameters, and $\theta$ is the parameter space of distribution of $G$. For more details about this family and its properties we refer readers to \cite{Alzaatreh2013a} and \cite{Alzaatreh2013b}. 

\item Weibull $G$ (\verb|weibullg|) family: The \verb|weibullg| is in fact the Weibull-X family of distributions introduced by Alzaatreh et al. (2013b). The cdf and pdf of the \verb|weibullg| family are given by
\begin{align}\label{weibullg}
F_{weibullg}(x,\Theta) &=1 - {e^{ - {{\left( {\frac{{ - \log \left( {1 - G(x,\theta )} \right)}}{b}} \right)}^a}}},\\
f_{weibullg}(x,\Theta) &= \frac{a}{{{b^a}}}\frac{{g(x,\theta )}}{{1 - G(x,\theta )}}{\left[ { - \log \left( {1 - G(x,\theta )} \right)} \right]^{a - 1}}{e^{ - {{\left( {\frac{{ - \log \left( {1 - G(x,\theta )} \right)}}{b}} \right)}^a}}},
\end{align}
where $\Theta=(a, b, \theta^{T})^{T}$ is the parameter space of the \verb|weibullg| family, $a>0$ and $b>0$ are the new induced shape parameters, and $\theta$ is the parameter space of distribution of $G$. Some works have been devoted to investigate the properties and applications of \verb|weibullg| family, see \cite{Nadarajah2016}.
\end{enumerate}
\section{MPS package: A guide to use in applications}
Cheng and Amin (1979, 1983) and independently Ranneby (1984) developed the maximum product of spacings (MPS) estimators. The MPS approach can be considered as an alternative to the maximum likelihood (ML) method for estimating the parameters of a continuous univariate distribution. Cheng and Amin (1979) proved the asymptotic property of the MPS estimators and proved that MPS estimators are as efficient as the ML estimators when they break down. Coolen and Newby (1991) proved that the MPS estimators have invariance property. For applications in statistical inference, we refer reader to Shah and Gokhale (1993) (for Burr XII Distributions), Fitzgerald (1996) (for generalized Pareto and log-logistic), Rahman and Pearson (2002) (for two-parameter exponential), Rahman and Pearson (2003) (for two-parameter Pareto), Wong and Li (2006) (for extreme value), Rahman et al. (2007) (for two-parameter gamma), Abouammoh and Alshingiti (2009), and Singh et al. (2014) (for generalized inverse exponential), and Singh et al. (2016) (for generalized inverse exponential under progressive type II censoring scheme) among them. Suppose $x_{(1)}, x_{(2)}, \dots, x_{(n)}$ are the ordered random observations of size $n$ drawn from a population with cdf $F(.,\theta)$ with unknown parameter space $\theta$. The MPS approach works on the basis of maximizing the mean of log-spacing function 
\begin{align*}\label{moran}
S(\theta)=\frac{1}{m}\sum_{i=1}^{m} \log \Bigl [ F\bigl(x_{(i)},\theta \bigr)-F\bigl(x_{(i-1)},\theta \bigr) \Bigr ],
\end{align*}
with respect to $\theta$ in which $m=n+1$, $F(x_{(0)},\theta)=0$ and $F(x_{(m)},\theta)=1$ with $m=n+1$. 
It can be shown that the Moran's statistic (S($\theta$) when $\theta$ is known) has asymptotic normal distribution. Also, a chi-square approximation exists for small samples whose mean and variance approximately are $m(\log(m)+0.57722)-0.5-1/(12m)$ and $m(\pi^2/6-1)-0.5-1/(6m)$, respectively, see \cite{Cheng1989}. Based on what mentioned above, the MPS approach is quite efficient in estimating the parameters of distributions with a shifted origin. So, hereafter we assume that all 24 $G$ families introduced in the previous section have an extra location parameter called $\mu$, and hence the cdf and pdf of the $G$ distribution are generally shown by $G(x,\theta^{*})$ and $g(x,\theta^{*})$, respectively, where $\theta^{*}=(\theta, \mu)^{T}$ in which $\theta$ is the parameter space of $G$ distribution.

The \verb|MPS| package has been developed for five tasks including: computing the cdf, computing the pdf, computing the quantile, generating random samples and estimating the parameters (using the MPS approach) of 24 $G$ families introduced in the previous section. For each of these $G$ families, distribution of $G$ is freely chosen from 15 standard distributions whose probability density functions, i.e., $g(x,\theta^{*})$ are given by the following.
\begin{itemize}
\item Birnbaum-Saunders (\verb|"birnbaum-saunders"|) with pdf
\begin{eqnarray}\label{birnbaumsaunderspdf}
g(x,\theta^{*}) = \frac{\sqrt{\frac{x-\mu}{\beta}}+\sqrt{\frac{\beta}{x-\mu}}}{2\alpha(x-\mu)}\phi\Biggl( \frac{\sqrt{\frac{x-\mu}{\beta}}-\sqrt{\frac{\beta}{x-\mu}}}{\alpha}\Biggr),
\end{eqnarray}
where $\phi(.)$ is the standard normal pdf, $x>\mu$ and $\theta^{*}=(\alpha, \beta, \mu)^{T}$ in which $\alpha> 0$, $\beta> 0$, $\mu \in \mathbb{R}$ are the shape, scale, and location parameters, respectively.

\item Burr XII (\verb|"burrxii"|) with pdf
\begin{eqnarray}\label{burrxiipdf}
g(x,\theta^{*}) = \alpha \beta (x-\mu)^{\beta-1}\bigl( 1+(x-\mu)^\beta \bigr) ^{-\alpha-1},
\end{eqnarray}
where $x>\mu$ and $\theta^{*}=(\alpha, \beta, \mu)^{T}$ in which $\alpha> 0$ and $\beta> 0$ are the first and second shape parameters and $\mu \in \mathbb{R}$ is location parameter.

\item Chen (\verb|"chen"|) with pdf
\begin{eqnarray}\label{chenpdf}
g(x,\theta^{*}) = \alpha \beta (x-\mu)^{\alpha-1}\exp \bigl((x-\mu)^\alpha \bigr)\exp \Bigl\{ -\beta \Bigl[ \exp \bigl((x-\mu)^\alpha \bigr) - 1 \Bigr]\Bigr\},
\end{eqnarray}
where $x>\mu$ and $\theta^{*}=(\alpha, \beta, \mu)^{T}$ in which $\alpha> 0$ and $\beta> 0$ are the first and second shape parameters and $\mu \in \mathbb{R}$ is location parameter.

\item
Chi-square (\verb|"chisq"|) with pdf
\begin{eqnarray}\label{chisqpdf}
g(x,\theta^{*}) = \Gamma^{-1}\left( \frac {\alpha}{2}\right) 2^{-\frac {\alpha}{2}} (x-\mu)^{\frac {\alpha}{2}-1}\exp\Bigl(-\frac {x-\mu}{2} \Bigr),
\end{eqnarray}
where $x>\mu$ and $\theta^{*}=(\alpha, \mu)^{T}$ in which $\alpha> 0$ and $\mu \in \mathbb{R}$ are degrees of freedom and location parameter, respectively.

\item
Exponential (\verb|"exp"|) with pdf
\begin{eqnarray}\label{exppdf}
g(x,\theta^{*}) =\alpha \exp \bigl(-\alpha (x-\mu)\bigr),
\end{eqnarray}
where $x>\mu$ and $\theta^{*}=(\alpha, \mu)^{T}$ in which $\alpha> 0$ and $\mu \in \mathbb{R}$ are the rate and location parameters, respectively.

\item
F (\verb|"f"|) with with pdf 
\begin{eqnarray}\label{fpdf}
g(x,\theta^{*}) =B^{-1}\Bigl(\frac{\alpha}{2},\frac{\beta}{2}\Bigr)\Bigl( \frac {\alpha}{\beta} \Bigr)^{\frac {\alpha}{2}} (x-\mu)^{\frac {\alpha}{2}-1}\Bigl(1 + \alpha\frac {x-\mu}{\beta} \Bigr)^{-\left(\frac {\alpha+\beta}{2} \right)},
\end{eqnarray}
where $x>\mu$ and $\theta^{*}=(\alpha, \beta, \mu)^{T}$ in which $\alpha> 0$ and $\beta> 0$ are the first and second degrees of freedom parameters and $\mu \in \mathbb{R}$ is location parameter.

\item
Frechet (\verb|"frechet"|) with pdf
\begin{eqnarray}\label{frechetpdf}
g(x,\theta^{*}) = \frac{\alpha}{ \beta} \Bigl( \frac {x-\mu}{\beta}\Bigr) ^{-\alpha-1}\exp\biggl\{ -\Bigl( \frac {x-\mu}{\beta}\Bigr)^{-\alpha} \biggr\},
\end{eqnarray}
where $x>\mu$ and $\theta^{*}=(\alpha, \beta, \mu)^{T}$ in which $\alpha>0$, $\beta>0$, and $\mu \in \mathbb{R}$ are the shape, scale, and location parameters, respectively.

\item
Gamma (\verb|"gamma"|) with pdf
\begin{eqnarray}\label{gammapdf}
g(x,\theta^{*}) = \bigl[ \beta^\alpha \Gamma(\alpha)\bigr]^{-1} (x-\mu)^{\alpha-1} \exp\Bigl( -\frac {x-\mu}{\beta}\Bigr),
\end{eqnarray}
where $x>\mu$ and $\theta^{*}=(\alpha, \beta, \mu)^{T}$ in which $\alpha>0$, $\beta>0$, and $\mu \in \mathbb{R}$ are the shape, scale, and location parameters, respectively.

\item
Gompertz (\verb|"gompertz"|) with pdf
\begin{eqnarray}\label{gompertzpdf}
g(x,\theta^{*}) = \alpha \exp \left\{ \beta (x-\mu) - \frac {\alpha}{\beta}\Bigl[ \exp \bigl(\beta (x-\mu)\bigr)-1\Bigr] \right\},
\end{eqnarray}
where $x>0$ and $\theta^{*}=(\alpha, \beta, \mu)^{T}$ in which $\alpha>0$, $\beta>0$, and $\mu \in \mathbb{R}$ are the first, second, and location parameters, respectively.

\item
Linear failure rate (\verb|"lfr"|) with pdf
\begin{eqnarray}\label{lfrpdf}
g(x,\theta^{*}) = \bigl(\alpha + \beta (x-\mu)\bigr)\exp\biggl\{ -\alpha x - \frac { \beta (x-\mu)^2}{2}\biggr\},
\end{eqnarray}
where $x>0$ and $\theta^{*}=(\alpha, \beta, \mu)^{T}$ in which $\alpha>0$, $\beta>0$, and $\mu \in \mathbb{R}$ are the first, second, and location parameters, respectively.

\item
Log-logistic (\verb|"log-logistic"|) with pdf
\begin{eqnarray*}\label{loglogisticpdf}
g(x,\theta^{*}) =\frac{ \alpha}{ \beta^{\alpha}} (x-\mu)^{\alpha-1} \left[ \Bigl( \frac {x-\mu}{\beta}\Bigr)^\alpha +1\right] ^{-2},
\end{eqnarray*}
where $x>\mu$ and $\theta^{*}=(\alpha, \beta, \mu)^{T}$ in which $\alpha>0$, $\beta>0$, and $\mu \in \mathbb{R}$ are the shape, scale, and location parameters, respectively.

\item
Log-normal (\verb|"log-normal"|) with pdf
\begin{eqnarray}\label{lognormalpdf}
g(x,\theta^{*}) = \bigl(\sqrt{2\pi} \beta (x-\mu) \bigr)^{-1}\exp\biggl\{ -\frac {1}{2}\left( \frac {\log(x-\mu) - \alpha}{\beta}\right) ^2\biggr\},
\end{eqnarray}
where $x>0$ and $\theta^{*}=(\alpha, \beta, \mu)^{T}$ in which $\alpha>0$ and $\beta>0$ are the first and second family parameters.

\item Lomax (\verb|"lomax"|) with pdf
\begin{eqnarray}\label{lomaxpdf}
g(x,\theta^{*}) = \alpha \beta \bigl( 1+\beta (x-\mu)\bigr)^{-(\alpha+1)},
\end{eqnarray}
where $x>\mu$ and $\theta^{*}=(\alpha, \beta, \mu)^{T}$ in which $\alpha>0$, $\beta>0$, $\mu \in \mathbb {R}$ are the shape, rate, and location parameters, respectively.

\item Rayleigh (\verb|"rayleigh"|) with pdf
\begin{eqnarray}\label{rayleighpdf}
g(x,\theta^{*}) = 2\frac {x-\mu}{\beta^2}\exp\biggl\{ -\Bigl( \frac {x-\mu}{\beta}\Bigr)^2 \biggr\},
\end{eqnarray}
where $x>\mu$ and $\theta^{*}=(\beta, \mu)^{T}$ in which $\beta>0$ and $\mu \in \mathbb {R}$ are the scale and location parameters, respectively.

\item Weibull (\verb|"weibull"|) with pdf
\begin{eqnarray}\label{weibullpdf}
g(x,\theta^{*}) = \frac {\alpha}{\beta} \Bigl( \frac {x-\mu}{\beta} \Bigr)^{\alpha - 1}\exp\biggl\{ -\Bigl( \frac {x-\mu}{\beta}\Bigr)^\alpha \biggr\},
\end{eqnarray}
where $x>\mu$ and $\theta^{*}=(\alpha, \beta, \mu)^{T}$ in which $\alpha>0$, $\beta>0$, $\mu \in \mathbb {R}$ are the shape, scale, and location parameters, respectively.
\end{itemize}

\subsection{R command for computing the pdf of $G$ families}
In this subsections, we give the general format of commands to compute the pdf of 24 $G$ families introduced in the Section 1 including \verb|betaexpg|, \verb|betag|, \verb|expexppg|, \verb|expg|, \verb|expgg|, \verb|expkumg|, \verb|gammag|, \verb|gammag1|, \verb|gammag2|, \verb|gbetag|, \verb|gexppg|, \verb|gmbetaexpg|, \verb|gtransg|, \verb|gxlogisticg|, \verb|kumg|, \verb|loggammag1|, \verb|loggammag2|, \verb|mbetag|, \verb|mog|, \verb|mokumg|, \verb|ologlogg|, \verb|texpsg|, \verb|weibullextg|, and \verb|weibullg|. The commands for computing the pdf are \verb|dbetaexpg(...)|, \verb|dbetag(...)|, \verb|dexpexppg(...)|, \verb|dexpg(...)|, \verb|dexpgg(...)|, \verb|dexpkumg(...)|, \verb|dgammag(...)|, \verb|dgammag1(...)|, \verb|dgammag2(...)|, \verb|dgbetag(...)|, \verb|dgexppg(...)|, \verb|dgmbetaexpg(...)|, \verb|dgtransg(...)|, \verb|dgxlogisticg(...)|, \verb|dkumg(...)|, \verb|dloggamm|\\ \verb|ag1(...)|, \verb|dloggammag2(...)|, \verb|dmbetag(...)|, \verb|dmog(...)|, \verb|dmokumg(...)|, \verb|dologlogg(...)|, \verb|dtexp|\\ \verb|sg(...)|, \verb|dweibullextg(...)|, and \verb|dweibullg(...)|, respectively. In the following, for instance, general format for computing the pdf of \verb|betaexpg| family and details about its arguments are given.
\begin{verbatim}
dbetaexpg(mydata, g, param, location = TRUE, log=FALSE)
\end{verbatim}
Details for command arguments are: 
\begin{itemize}
\item \verb|mydata| : Vector of observations. 
\item \verb|g| : The name of family's pdf including: \verb|"birnbaum-saunders"|, \verb|"burrxii"|, \verb|"chisq"|, \verb|"chen"|, \verb|"exp"|, \verb|"f"|, \verb|"frechet"|, \verb|"gamma"|, \verb|"gompertz"|, \verb|"lfr"|, \verb|"log-normal"|, \verb|"log-logistic"|, \verb|"lomax"|, \verb|"rayleigh"|, and \verb|"weibull"|.

\item \verb|param| : The parameter space can be of the form $\Theta=\bigl(a, {\theta^{*}}^{T}\bigr)^{T}$, $\Theta=\bigl(a, b, {\theta^{*}}^{T}\bigr)^{T}$, or $\Theta=\bigl(a, b, d, {\theta^{*}}^{T}\bigr)^{T}$, where ${\theta^{*}}$ is the parameter space of shifted $G$ distribution as mentioned before. The general form for $\theta^{*}$ can be ${\theta^{*}}=(\alpha, \mu)^{T}$, ${\theta^{*}}=(\alpha, \beta, \mu)^{T}$, or ${\theta^{*}}=(\beta, \mu)^{T}$. As it is seen, the location parameter is placed in the last component of ${\theta}^{*}$. The induced parameters $a$, $b$, and $d$ are listed before ${\theta^{*}}^{T}$ in parameter 
space ${\Theta^{}}$.
\item \verb|location| : If \verb|FALSE|, then the location parameter is absent.
\item \verb|log| : If \verb|TRUE|, then the logarithm of pdf is returned.
\end{itemize}

\subsection{R command for computing the cdf of $G$ families}
In this subsections, we give the general format of commands to compute the cdf of 24 $G$ families introduced in the Section 1 including \verb|betaexpg|, \verb|betag|, \verb|expexppg|, \verb|expg|, \verb|expgg|, \verb|expkumg|, \verb|gammag|, \verb|gammag1|, \verb|gammag2|, \verb|gbetag|, \verb|gexppg|, \verb|gmbetaexpg|, \verb|gtransg|, \verb|gxlogisticg|, \verb|kumg|, \verb|loggammag1|, \verb|loggammag2|, \verb|mbetag|, \verb|mog|, \verb|mokumg|, \verb|ologlogg|, \verb|texpsg|, \verb|weibullextg|, and \verb|weibullg|. The commands for computing the cdf are \verb|pbetaexpg(...)|, \verb|pbetag(...)|, \verb|pexpexppg(...)|, \verb|pexpg(...)|, \verb|pexpgg(...)|, \verb|pexpkumg(...)|, \verb|pgammag(...)|, \verb|pgammag1(...)|, \verb|pgammag2(...)|, \verb|pgbetag(...)|, \verb|pgexppg(...)|, \verb|pgmbetaexpg(...)|, \verb|pgtransg(...)|, \verb|pgxlogisticg(...)|, \verb|pkumg(...)|, \verb|ploggamm|\\ \verb|ag1(...)|, \verb|ploggammag2(...)|, \verb|pmbetag(...)|, \verb|pmog(...)|, \verb|pmokumg(...)|, \verb|pologlogg(...)|, \verb|ptexp|\\ \verb|sg(...)|, \verb|pweibullextg(...)|, and \verb|pweibullg(...)|, respectively. In the following, for instance, general format for computing the cdf of \verb|betaexpg| family and details about its arguments are given.
\begin{verbatim}
pbetaexpg(mydata, g, param, location = TRUE, log.p = FALSE, lower.tail = TRUE)
\end{verbatim}
Details for command arguments are: 
\begin{itemize}
\item \verb|mydata| : Vector of observations.
\item \verb|g| : The name of family's pdf including: \verb|"birnbaum-saunders"|, \verb|"burrxii"|, \verb|"chisq"|, \verb|"chen"|, \verb|"exp"|, \verb|"f"|, \verb|"frechet"|, \verb|"gamma"|, \verb|"gompertz"|, \verb|"lfr"|, \verb|"log-normal"|, \verb|"log-logistic"|, \verb|"lomax"|, \verb|"rayleigh"|, and \verb|"weibull"|.
\item \verb|param| : The parameter space can be of the form $\Theta=\bigl(a, {\theta^{*}}^{T}\bigr)^{T}$, $\Theta=\bigl(a, b, {\theta^{*}}^{T}\bigr)^{T}$, or $\Theta=\bigl(a, b, d, {\theta^{*}}^{T}\bigr)^{T}$, where ${\theta^{*}}$ is the parameter space of shifted $G$ distribution as mentioned before. The general form for $\theta^{*}$ can be ${\theta^{*}}=(\alpha, \mu)^{T}$, ${\theta^{*}}=(\alpha, \beta, \mu)^{T}$, or ${\theta^{*}}=(\beta, \mu)^{T}$. As it is seen, the location parameter is placed in the last component of ${\theta}^{*}$. The induced parameters $a$, $b$, and $d$ are listed before ${\theta^{*}}^{T}$ in parameter 
space ${\Theta^{}}$.
\item \verb|location| : If \verb|FALSE|, then the location parameter is absent.
\item \verb|log.p| : If \verb|TRUE|, then the logarithm of cdf is returned and quantile is computed for $\exp(-p)$.
\item \verb|lower.tail| : If \verb|FALSE|, then 1-cdf is returned and quantile is computed for $1-p$.
\end{itemize}

\subsection{R command for computing the quantile of $G$ families}
Here, we give the general format of commands to compute the quantiles of 24 $G$ families introduced in the Section 1 including \verb|betaexpg|, \verb|betag|, \verb|expexppg|, \verb|expg|, \verb|expgg|, \verb|expkumg|, \verb|gammag|, \verb|gammag1|, \verb|gammag2|, \verb|gbetag|, \verb|gexppg|, \verb|gmbetaexpg|, \verb|gtransg|, \verb|gxlogisticg|, \verb|kumg|, \verb|loggammag1|, \verb|loggammag2|, \verb|mbetag|, \verb|mog|, \verb|mokumg|, \verb|ologlogg|, \verb|texpsg|, \verb|weibullextg|, and \verb|weibullg|. The commands for computing the quantile are \verb|qbetaexpg(...)|, \verb|qbetag(...)|, \verb|qexpexppg(...)|, \verb|qexpg(...)|, \verb|qexpgg(...)|, \verb|qexpkumg(...)|, \verb|qgammag(...)|, \verb|qgammag1(...)|, \verb|qgammag2(...)|, \verb|qgbetag(...)|, \verb|qgexppg(...)|, \verb|qgmbetaexpg(...)|, \verb|qgtransg(...)|, \verb|qgxlogisticg(...)|, \verb|qkumg(...)|, \verb|qloggammag1(...)|, \verb|qlog|\\ \verb|gammag2(...)|, \verb|qmbetag(...)|, \verb|qmog(...)|, \verb|qmokumg(...)|, \verb|qologlogg(...)|, \verb|qtexpsg(...)|, \verb|qweib|\\ \verb|ullextg(...)|, and \verb|qweibullg(...)|, respectively. In the following, for instance, general format for computing the quantile of \verb|betaexpg| family and details about its arguments are given.
\begin{verbatim}
qbetaexpg(p, g, param, location = TRUE, log.p = FALSE, lower.tail = TRUE)
\end{verbatim}
Details for command arguments are: 
\begin{itemize}
\item \verb|p| : A vector of value(s) between 0 and 1 at which the quantile needs to be computed at those points.
\item \verb|g| : The name of family's pdf including: \verb|"birnbaum-saunders"|, \verb|"burrxii"|, \verb|"chisq"|, \verb|"chen"|, \verb|"exp"|, \verb|"f"|, \verb|"frechet"|, \verb|"gamma"|, \verb|"gompertz"|, \verb|"lfr"|, \verb|"log-normal"|, \verb|"log-logistic"|, \verb|"lomax"|, \verb|"rayleigh"|, and \verb|"weibull"|.
\item \verb|param| : The parameter space can be of the form $\Theta=\bigl(a, {\theta^{*}}^{T}\bigr)^{T}$, $\Theta=\bigl(a, b, {\theta^{*}}^{T}\bigr)^{T}$, or $\Theta=\bigl(a, b, d, {\theta^{*}}^{T}\bigr)^{T}$, where ${\theta^{*}}$ is the parameter space of shifted $G$ distribution as mentioned before. The general form for $\theta^{*}$ can be ${\theta^{*}}=(\alpha, \mu)^{T}$, ${\theta^{*}}=(\alpha, \beta, \mu)^{T}$, or ${\theta^{*}}=(\beta, \mu)^{T}$. As it is seen, the location parameter is placed in the last component of ${\theta}^{*}$. The induced parameters $a$, $b$, and $d$ are listed before ${\theta^{*}}^{T}$ in parameter 
space ${\Theta^{}}$.
\item \verb|location| : If \verb|FALSE|, then the location parameter is absent.
\item \verb|log.p| : If \verb|TRUE|, then the logarithm of cdf is returned and quantile is computed for $\exp(-p)$.
\item \verb|lower.tail| : If \verb|FALSE|, then 1-cdf is returned and quantile is computed for $1-p$.
\end{itemize}
\subsection{R command for simulating random generation from $G$ families}\label{simul}
Here, we give the general format of commands for simulating realizations from each of 24 $G$ families introduced in the Section 1. These include \verb|betaexpg|, \verb|betag|, \verb|expexppg|, \verb|expg|, \verb|expgg|, \verb|expkumg|, \verb|gammag|, \verb|gammag1|, \verb|gammag2|, \verb|gbetag|, \verb|gexppg|, \verb|gmbetaexpg|, \verb|gtransg|, \verb|gxlogisticg|, \verb|kumg|, \verb|loggammag1|, \verb|loggammag2|, \verb|mbetag|, \verb|mog|, \verb|mokumg|, \verb|ologlogg|, \verb|texpsg|, \verb|weibullextg|, and \verb|weibullg|. The commands for generating realizations are: \verb|rbetaexpg(...)|, \verb|rbetag(...)|, \verb|rexpexppg(...)|, \verb|rexpg(...)|, \verb|rexpgg(...)|, \verb|rexpkumg(...)|, \verb|rgammag(...)|, \verb|rgammag1(...)|, \verb|rgammag2(...)|, \verb|rgbe|\\
\verb|tag(...)|, \verb|rgexppg(...)|, \verb|rgmbetaexpg(...)|, \verb|rgtransg(...)|, \verb|rgxlogisticg(...)|, \verb|rkumg(...)|, \verb|rloggammag1(...)|, \verb|rloggammag2(...)|, \verb|rmbetag(...)|, \verb|rmog(...)|, \verb|rmokumg(...)|, \verb|rologlogg(..|\\\verb|.)|, \verb|rtexpsg(...)|, \verb|rweibullextg(...)|, and \verb|rweibullg(...)|, respectively. In the following, for instance, general format for simulating realizations from \verb|betaexpg| family and details about its arguments are given.
\begin{verbatim}
rbetaexpg(n, g, param, location = TRUE)
\end{verbatim}
Details for command arguments are: 
\begin{itemize}
\item \verb|n| : The number of realizations needed for generation.
\item \verb|g| : The name of family's pdf including: \verb|"birnbaum-saunders"|, \verb|"burrxii"|, \verb|"chisq"|, \verb|"chen"|, \verb|"exp"|, \verb|"f"|, \verb|"frechet"|, \verb|"gamma"|, \verb|"gompertz"|, \verb|"lfr"|, \verb|"log-normal"|, \verb|"log-logistic"|, \verb|"lomax"|, \verb|"rayleigh"|, and \verb|"weibull"|.
\item \verb|param| : The parameter space can be of the form $\Theta=\bigl(a, {\theta^{*}}^{T}\bigr)^{T}$, $\Theta=\bigl(a, b, {\theta^{*}}^{T}\bigr)^{T}$, or $\Theta=\bigl(a, b, d, {\theta^{*}}^{T}\bigr)^{T}$, where ${\theta^{*}}$ is the parameter space of shifted $G$ distribution as mentioned before. The general form for $\theta^{*}$ can be ${\theta^{*}}=(\alpha, \mu)^{T}$, ${\theta^{*}}=(\alpha, \beta, \mu)^{T}$, or ${\theta^{*}}=(\beta, \mu)^{T}$. As it is seen, the location parameter is placed in the last component of ${\theta}^{*}$. The induced parameters $a$, $b$, and $d$ are listed before ${\theta^{*}}^{T}$ in parameter 
space ${\Theta^{}}$.
\item \verb|location| : If \verb|FALSE|, then the location parameter is absent.
\end{itemize}

\subsection{R command for estimating the parameters of the $G$ families}
In this subsections we give the general format of commands for estimating the parameters of 24 $G$ families introduced in the Section 1. These include \verb|betaexpg|, \verb|betag|, \verb|expexppg|, \verb|expg|, \verb|expgg|, \verb|expkumg|, \verb|gammag|, \verb|gammag1|, \verb|gammag2|, \verb|gbetag|, \verb|gexppg|, \verb|gmbetaexpg|, \verb|gtransg|, \verb|gxlogisticg|, \verb|kumg|, \verb|loggammag1|, \verb|loggammag2|, \verb|mbetag|, \verb|mog|, \verb|mokumg|, \verb|ologlogg|, \verb|texpsg|, \verb|weibullextg|, and \verb|weibullg|. The commands for estimating the parameters are: \verb|mpsbetaexpg(...)|, \verb|mpsbetag(...)|, \verb|mpsexpexppg(...)|, \verb|mpsexpg(...)|, \verb|mpsexpgg(...)|, \verb|mpsexpkumg(...)|, \verb|mpsgammag(...)|, \verb|mpsgam|\\
\verb|mag1(...)|, \verb|mpsgammag2(...)|, \verb|mpsgbetag(...)|, \verb|mpsgexppg(...)|, \verb|mpsgmbetaexpg(...)|, \verb|mpsgtra|\\
\verb|nsg(...)|, \verb|mpsgxlogisticg(...)|, \verb|mpskumg(...)|, \verb|mpsloggammag1(...)|, \verb|mpsloggammag2(...)|, \verb|mpsmbetag(...)|, \verb|mpsmog(...)|, \verb|mpsmokumg(...)|, \verb|mpsologlogg(...)|, \verb|mpstexpsg(...)|, \verb|mpsweibu|\\
\verb|llextg(...)|, and \verb|mpsweibullg(...)|, respectively. In the following, for instance, general format for estimating the parameters of \verb|betaexpg| family and details about its arguments are given.
\begin{verbatim}
mpsbetaexpg(mydata, g, location = TRUE, method, sig.level)
\end{verbatim}
Details for command arguments are: 
\begin{itemize}
\item \verb|mydata| : Vector of observations.
\item \verb|g| : The name of family's pdf including: \verb|"birnbaum-saunders"|, \verb|"burrxii"|, \verb|"chisq"|, \verb|"chen"|, \verb|"exp"|, \verb|"f"|, \verb|"frechet"|, \verb|"gamma"|, \verb|"gompertz"|, \verb|"lfr"|, \verb|"log-normal"|, \verb|"log-logistic"|, \verb|"lomax"|, \verb|"rayleigh"|, and \verb|"weibull"|.
\item \verb|location| : If \verb|FALSE|, then the location parameter is absent.
\item \verb|method| : The used method for maximizing the sum of log-spacing function. It will be \verb|"BFGS"|, \verb|"CG"|, \verb|"L-BFGS-B"|, \verb|"Nelder-Mead"|, or \verb|"SANN"|.
\item \verb|sig.level| : Significance level for the approximated chi-square goodness-of-fit test.
\end{itemize}
The details of output of \verb|mpsbetaexpg(mydata, g, location = TRUE, method, sig.level)| are:
\begin{itemize}
\item Estimated parameter space $\widehat{\Theta}$, represented by \verb|$MPS|. 
\item A sequence of goodness-of-fit statistics such as: Akaike Information Criterion (\verb|AIC|), Consistent Akaike Information Criterion (\verb|CAIC|), Bayesian Information Criterion (\verb|BIC|), Hannan-Quinn information criterion (\verb|HQIC|), Cramer-von Misses statistic (\verb|CM|), Anderson Darling statistic (\verb|AD|), log-likelihood statistic (\verb|log|), and Moran's statistic (\verb|Moran|). These measures are represented by \verb|$Measures|
\item Kolmogorov-Smirnov test statistic and corresponding \verb|p-value|, represented by \verb|$KS|.
\item Chi-square test statistic, critical upper tail chi-square distribution, related \verb|p-value|, represented by \verb|$chi-square|.
\item Convergence status, represented by \verb|$Convergence Status|.
\end{itemize}
We note that the package is available from the Comprehensive R Archive Network (CRAN) at \url{https://cran.r-project.org/package=MPS}.
\section{Examples and illustrations}
Here, we provide some examples and real data applications to check the performance of the \verb|MPS| package. Firstly, we compute the cdf and pdf of the \verb|betaexpg| family when $G$ is three-parameter gamma distribution. Secondly, we use the \verb|MPS| package to estimate the parameters of \verb|weibullg|, \verb|kumg|, and \verb|mog| families when these families are applied to the three sets of real data. Finally, the mechanism of random number generation will be checked for \verb|loggammag1| family when $G$ is supposed to be \verb|"birnbaum-saunders"|, \verb|"log-logistic"|, \verb|"lomax"|, and \verb|"weibull"|.
\subsection{Computing the cdf and pdf}
\begin{figure}[h!]
\centering
\begin{subfigure}[b]{0.45\linewidth}
\includegraphics[width=\linewidth]{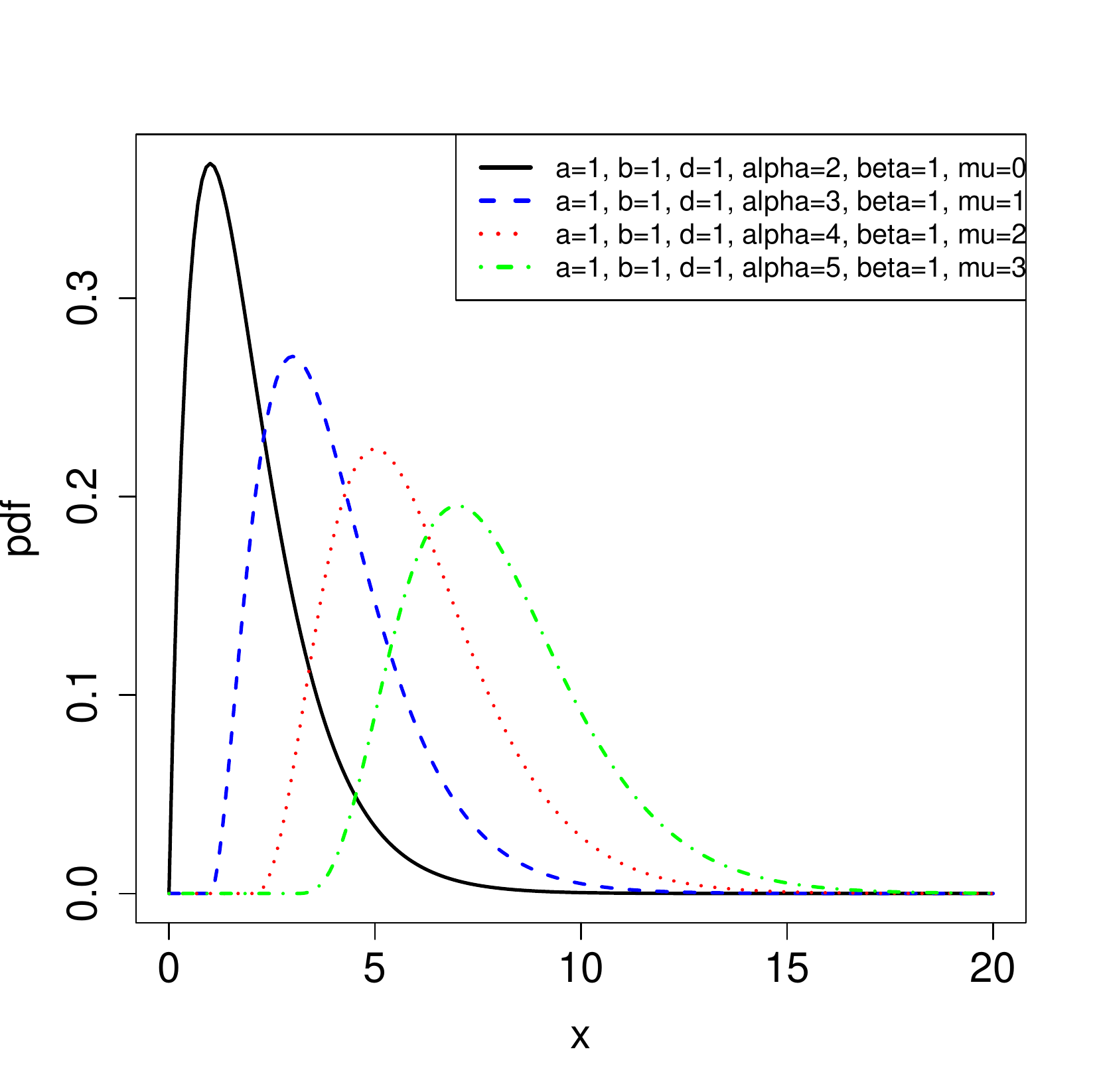}
\caption{pdf plot}
\end{subfigure}
\begin{subfigure}[b]{0.45\linewidth}
\includegraphics[width=\linewidth]{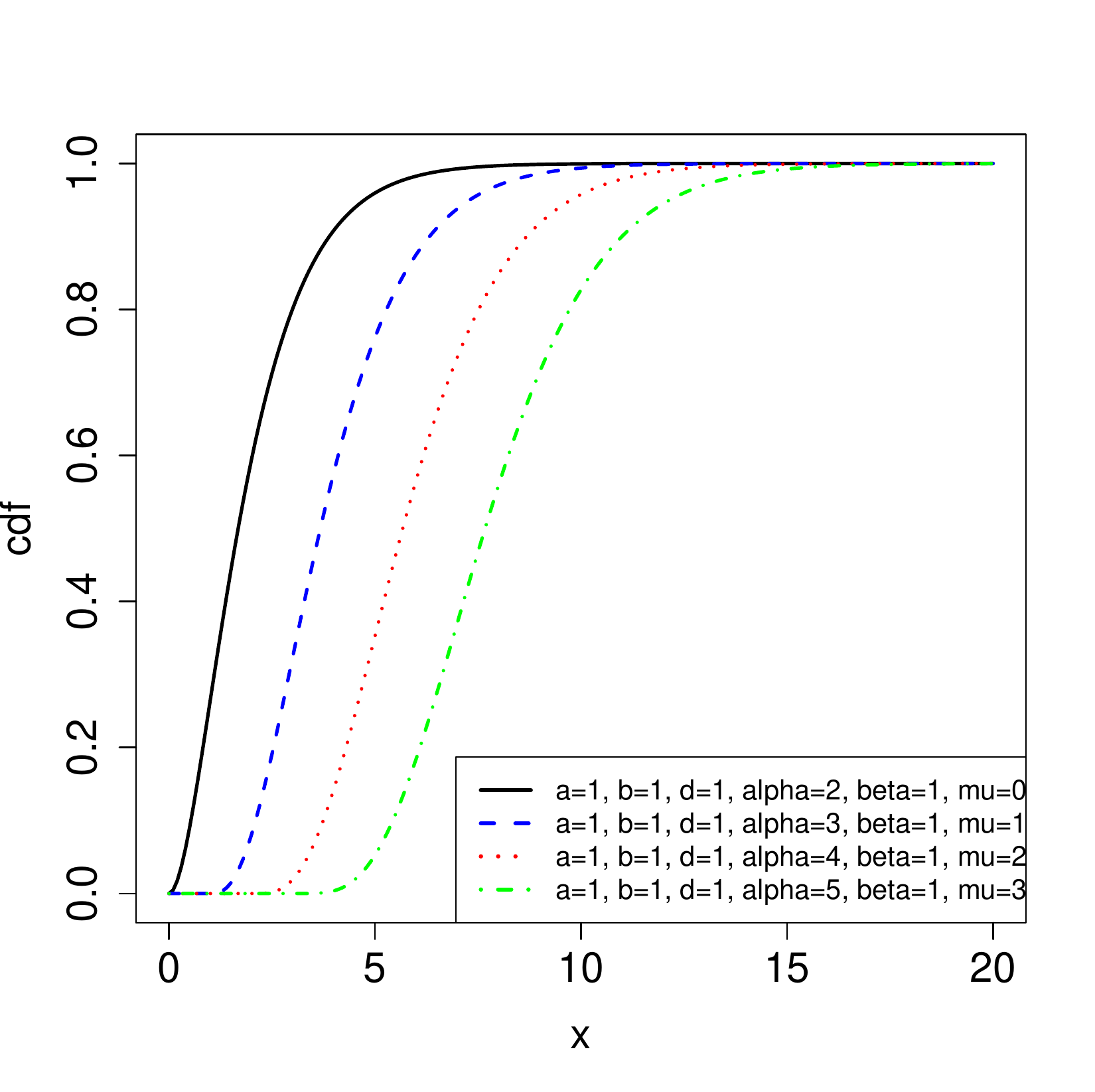}
\caption{cdf plot}
\end{subfigure}
\caption{Plots of pdf and cdf for four members of \texttt{betaexpg} family when G is three-parameter gamma distribution with shape, scale, and location parameters as \texttt{alpha}, \texttt{beta}, and \texttt{mu}, respectively.}
\label{fig1}
\end{figure}

The following commands will produce the pdf plot of four members of \verb|betaexpg| family when $G$ has distribution with pdf given in (\ref{gammapdf}). The results are displayed in left-hand side of Figure \ref{fig1}. 
\begin{verbatim}
R> x <- seq(0, 20, 0.1)
R> y1 <- dbetaexpg(x, "gamma", c(1,1,1,2,1,0))
R> y2 <- dbetaexpg(x, "gamma", c(1,1,1,3,1,1))
R> y3 <- dbetaexpg(x, "gamma", c(1,1,1,4,1,2))
R> y4 <- dbetaexpg(x, "gamma", c(1,1,1,5,1,3))
R> xrange <- range(x)
R> yrange <- range(y1, y2, y3, y4)
R> plot(x, y1, type="l", xlab="x", ylab="pdf", xlim=xrange, ylim=yrange, lty=1,
R> lines(x, y2, col = "blue", lty=2)
R> lines(x, y3, col = "red", lty=3)
R> lines(x, y4, col = "green", lty=4)
R> cols<-c("black","blue","red","green")
R> legend(7.5, 0.4, legend=c("a=1, b=1, d=1, alpha=2, beta=1, mu=0",
+ "a=1, b=1, d=1, alpha=3, beta=1, mu=1",
+ "a=1, b=1, d=1, alpha=4, beta=1, mu=2",
+ "a=1, b=1, d=1, alpha=5, beta=1, mu=3"), col=cols, lty=1:4, lwd=2.5, cex=1)
\end{verbatim}
The following commands will produce the cdf plot of four members of \verb|betaexpg| family when $G$ has distribution with pdf given in (\ref{gammapdf}). The results are displayed in right-hand side of Figure \ref{fig1}. 
\begin{verbatim}
R> x <- seq(0, 20, 0.1)
R> y1 <- pbetaexpg(x, "gamma", c(1,1,1,2,1,0))
R> y2 <- pbetaexpg(x, "gamma", c(1,1,1,3,1,1))
R> y3 <- pbetaexpg(x, "gamma", c(1,1,1,4,1,2))
R> y4 <- pbetaexpg(x, "gamma", c(1,1,1,5,1,3))
R> xrange <- range(x)
R> yrange <- range(y1, y2, y3, y4)
R> plot(x, y1, type="l", xlab="x", ylab="cdf", xlim=xrange, ylim=yrange, lty=1)
R> lines(x, y2, col = "blue", lty=2)
R> lines(x, y3, col = "red", lty=3)
R> lines(x, y4, col = "green", lty=4)
R> cols<-c("black","blue","red","green")
R> legend(7.5, 0.4, legend=c("a=1, b=1, d=1, alpha=2, beta=1, mu=0",
+ "a=1, b=1, d=1, alpha=3, beta=1, mu=1",
+ "a=1, b=1, d=1, alpha=4, beta=1, mu=2",
+ "a=1, b=1, d=1, alpha=5, beta=1, mu=3"), col=cols, lty=1:4, lwd=2.5, cex=1)
\end{verbatim}

\subsection{Estimating the parameters}
The performance of the \verb|MPS| package is demonstrated by analyzing three sets of real data. For the first set the usual ML estimators are not found while, the MPS counterparts exist and \verb|MPS| package find them. Two other applications verify that presence of the location parameter in the fitted model makes it more appropriate.
\par As the first real data application, we focus on fatigue life in hours of 10 bearings that initially reported by McCool (1974) analyzed by several researchers. For modelling fatigue life of bearings (denoted here as bearing) data via \verb|weibullg| family, we use the following commands.
\begin{verbatim}
R>x<-c(152.7, 172.0, 172.5, 173.3, 193.0, 204.7, 216.5, 234.9, 262.6, 422.6)
R>mpsweibullg(x,"weibull",TRUE,"Nedler-Mead",0.05)
\end{verbatim}
The output is
\begin{verbatim}
$MPS
[1]   0.9988519   0.9708349   0.8618143  83.4125577 147.1825435

$Measures
      AIC     CAIC      BIC     HQIC         CM        AD       log    Moran
 116.7875 131.7875 118.3004 115.1278 0.06325825 0.3809677 -53.39375 31.37394

$KS
 statistic   p-value
 0.2042573 0.7266451

$`chi-square`
 statistic chi-value  p-value
  12.88606  18.30704 0.230112

$`Convergence Status`
     [,1]                 
[1,] "Algorithm Converged"
\end{verbatim}
The estimated $\widehat{\Theta}$ for \verb|weibullg| family (with cdf given in (\ref{weibullg}) in which $G$ has a distribution with pdf given in (\ref{weibullpdf})) is $\widehat{\Theta}=(0.998851, 0.970834, 0.861814, 83.412557, 147.182543)^{T}$ where $\hat{a}=0.998851$ and $\hat{b}=0.970834$ are induced shape parameters and $\widehat{\theta^{*}}=(0.861814, 83.412557, 147.182543)^{T}$ is the estimated parameter space of three-parameter Weibull distribution with $\hat{\alpha}=0.998851$, $\hat{\beta}=0.970834$, and $\hat{\mu}=147.182543$. Other features of the above output are given by the following. The Akaike information criterion (\texttt{116.7875}), consistent Akaike information criterion (\texttt{131.7875}), Bayesian information criterion (\texttt{118.3004}), Hannan-Quinn information criterion (\texttt{115.1278}), Cramer-von Misses statistic (\texttt{0.06325825}), Anderson Darling statistic (\texttt{0.3809677}), log-likelihood statistic (\texttt{-53.39375}), Moran's statistic (\texttt{31.37394}), Kolmogorov-Smirnov test statistic (\texttt{0.2042573}), corresponding p-value (\texttt{0.7266451}), the chi-square test statistic (\texttt{12.88606}), critical upper tail chi-square distribution (\texttt{18.30704}), related p-value (\texttt{0.230112}), and convergence status (\texttt{"Algorithm Converged"}). For bearing data, as pointed out by Nagatsuka et al. (2013), the usual ML estimators break down. The estimated induced shape parameters are close to one ($\hat{a}=0.998851$ and $\hat{b}=0.970834$) that means a three-parameter Weibull distribution with shape, scale, and location parameters given, respectively, by 0.861814, 83.412557, and 147.182543 is an appropriate model for bearing data. The latter can be shown using a likelihood ratio test.
\par As the second real data application, we consider the large recorded intensities (in Richter scale) of the earthquake in seismometer locations in western North America between 1940 and 1980. The related features were reported by \cite{Joyner1981}. Among the features, we focus on the 182 distances from the seismological measuring station to the epicenter of the earthquake (in km) as the variable of interest. We apply the \verb|kumg| family with cdf given in (\ref{kumg}) to the large recorded intensities of the earthquake (denoted here as earthquake) data in two cases including: 1- when $G$ follows a three-parameter Birnbaum-Saunders distribution with pdf given in (\ref{birnbaumsaunderspdf}), and 2- $G$ follows a Birnbaum-Saunders distribution. For this, we use the following commands.
\begin{verbatim}
R>x<-c(7.5,8.8,8.9,9.4,9.7,9.7,10.5,10.5,12,12.2,12.8,14.6,14.9,17.6,23.9,25,2.9,
+    3.2,7.6,17,8,10,10,8,19,21,13,22,29,31,5.8,12,12.1,20.5,20.5,25.3,35.9,36.1,
+    36.3,38.5,41.4,43.6,44.4,46.1,47.1,47.7,49.2,53.1,4,10.1,11.1,17.7,22.5,26.5,
+    29,30.9,37.8,48.3,62,50,16,62,1.2,1.6,9.1,3.7,5.3,7.4,17.9,19.2,23.4,30,38.9,
+    10.8,15.7,16.7,20.8,28.5,33.1,40.3,8,32,30,31,16.1,63.6,6.6,9.3,13,17.3,105,
+    112,123,5,23.5,26,0.5,0.6,1.3,1.4,2.6,3.8,4,5.1,6.2,6.8,7.5,7.6,8.4,8.5,8.5,
+    10.6,12.6,12.7,12.9,14,15,16,17.7,18,22,22,23,23.2,29,32,32.7,36,43.5,49,60,
+    64,105,122,141,200,45,130,147,187,197,203,211,17,19.6,20.2,21.1,21.9,66,87,
+    23.4,24.6,25.7,28.6,37.4,46.7,56.9,60.7,61.4,62,64,82,88,91,12,24.2,148,42,
+    85,107,109,156,224,293,359,370,25.4,32.9,92.2,45,145,300)
R> mpskumg(x,"birnbaum-saunders",TRUE,"BFGS",0.05)
$MPS
[1]   3.419681  35.382782   5.180796 222.382191  -3.728349

$Measures
      AIC     CAIC      BIC     HQIC         CM        AD       log    Moran
 1737.207 1737.548 1753.227 1743.701 0.03211737 0.2332114 -863.6034 954.8089

$KS
  statistic   p-value
 0.03819923 0.9532803

$`chi-square`
 statistic chi-value p-value
  3.972337  214.4771       1

$`Convergence Status`
     [,1]                 
[1,] "Algorithm Converged"

R> mpskumg(x,"birnbaum-saunders",FALSE,"BFGS",0.05)
$MPS
[1] 2.2185211 0.3324036 1.7054161 3.2352482

$Measures
      AIC    CAIC     BIC    HQIC        CM       AD       log    Moran
 1754.054 1754.28 1766.87 1759.25 0.5519759 2.879096 -873.0271 967.3399

$KS
 statistic    p-value
 0.1044829 0.03760838

$`chi-square`
 statistic chi-value p-value
  25.14594  214.4771       1

$`Convergence Status`
     [,1]                 
[1,] "Algorithm Converged"
\end{verbatim}
\begin{figure}[h!]
\centering
\begin{subfigure}[b]{0.45\linewidth}
\includegraphics[width=\linewidth]{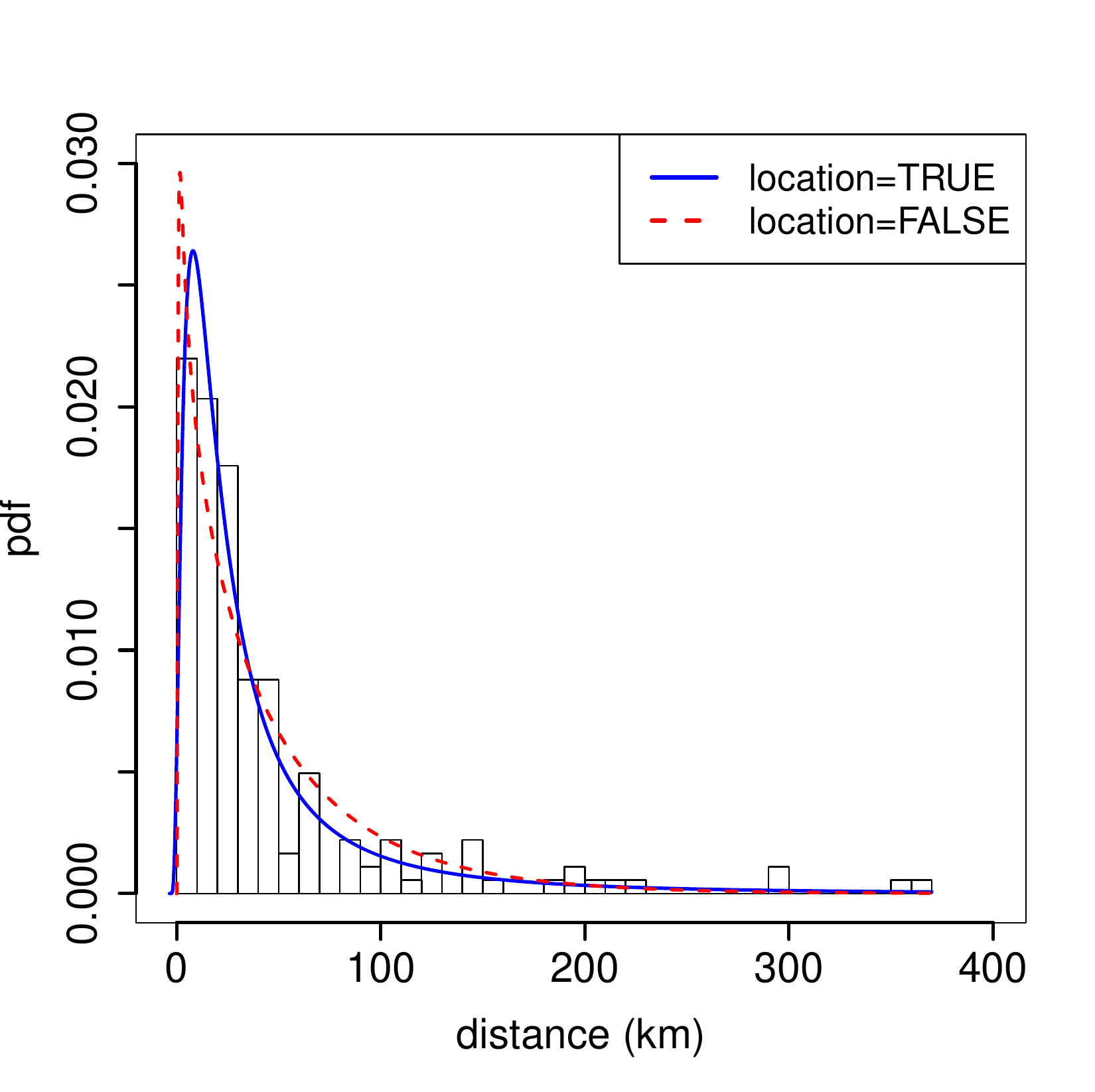}
\caption{pdf plot}
\end{subfigure}
\begin{subfigure}[b]{0.45\linewidth}
\includegraphics[width=\linewidth]{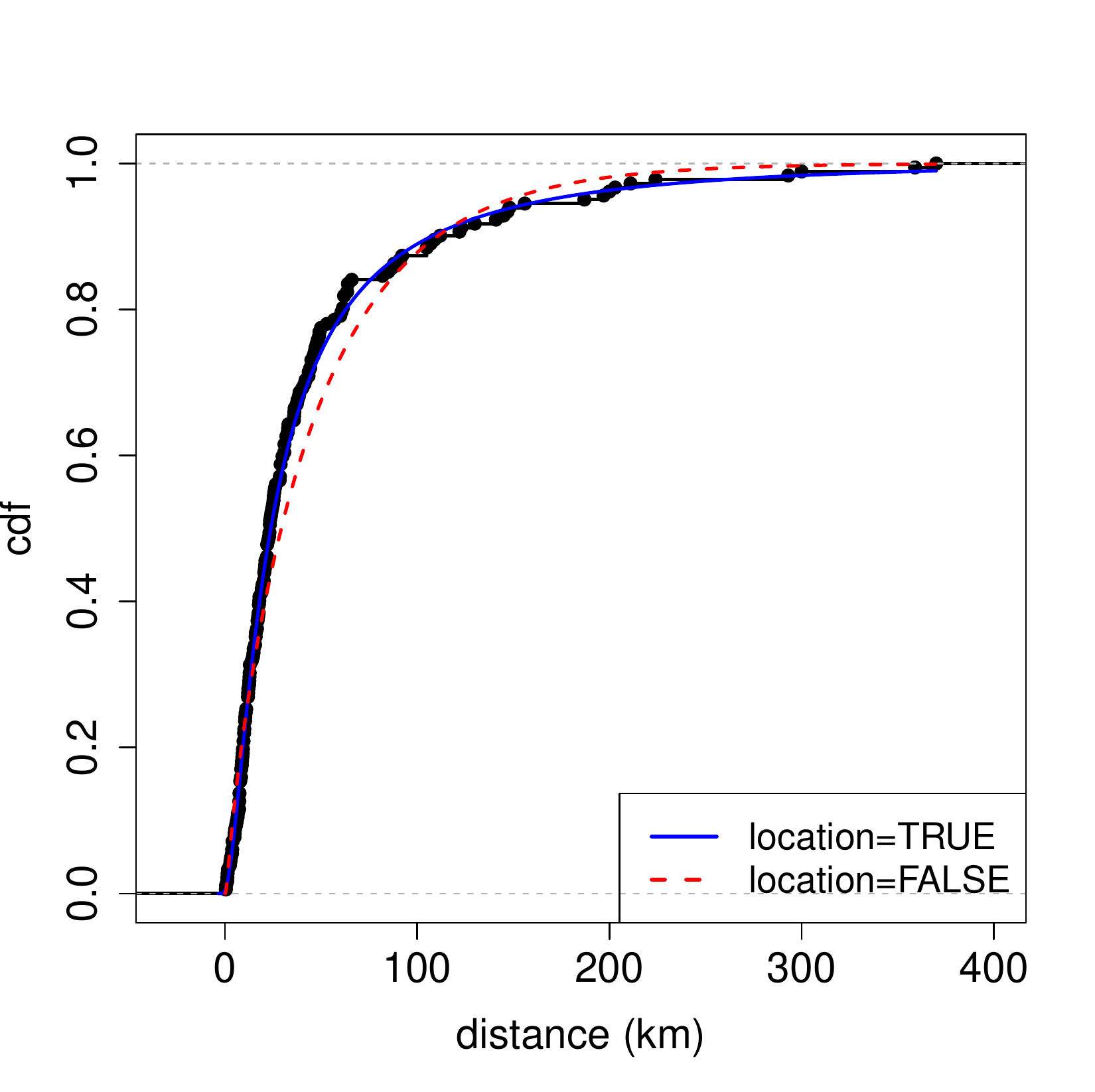}
\caption{cdf plot}
\end{subfigure}
\caption{Histogram and fitted probability density functions for earthquake data are displayed in left hand-side subfigure. The empirical distribution function and fitted cumulative distribution functions for earthquake data are displayed is the right hand-side subfigure. In each subfigure two cases are considered. Those include the presence of the location parameter (location=TRUE) and absence of the location parameter (location=FALSE).}
\label{fig2}
\end{figure}
When the location is present, we see $\widehat{\Theta}=(3.419681, 35.382782, 5.180796, 222.382191, -3.728349)^{T}$ in which $\hat{a}=3.419681$ and $\hat{b}=35.382782$ are induced shape parameters. The estimated parameter space of three-parameter Birnbaum-Saunders distribution is $\widehat{\theta^{*}}=(5.180796, 222.382191, -3.728349)^{T}$ with $\hat{\alpha}=5.180796$, $\hat{\beta}=222.382191$, and $\hat{\mu}=-3.728349$. Also, when the location is absent, we have $\widehat{\Theta}=(2.2185211, 0.3324036, 1.7054161, 3.2352482)^{T}$ in which $\hat{a}=2.2185211$ and $\hat{b}=0.3324036$ are induced shape parameters and $\widehat{\theta^{*}}=(1.7054161, 3.2352482)^{T}$ is estimated parameter space of two-parameter Birnbaum-Saunders distribution in which $\hat{\alpha}=1.7054161$ and $\hat{\beta}=3.2352482$. Based on above output, in the absence of location parameter, the Kolmogorov-Smirnov test statistic (\texttt{0.1044829}) and corresponding p-value (\texttt{0.03760838}) suggest that the \verb|kumg| family is not appropriate model, while in the presence of location parameter, the \verb|kumg| family is accepted. As Figure \ref{fig2} shows, when the location parameter is present, the fitted cdf captures well the general shape of the empirical distribution function. 
\par
\begin{figure}[h!]
\centering
\begin{subfigure}[b]{0.45\linewidth}
\includegraphics[width=\linewidth]{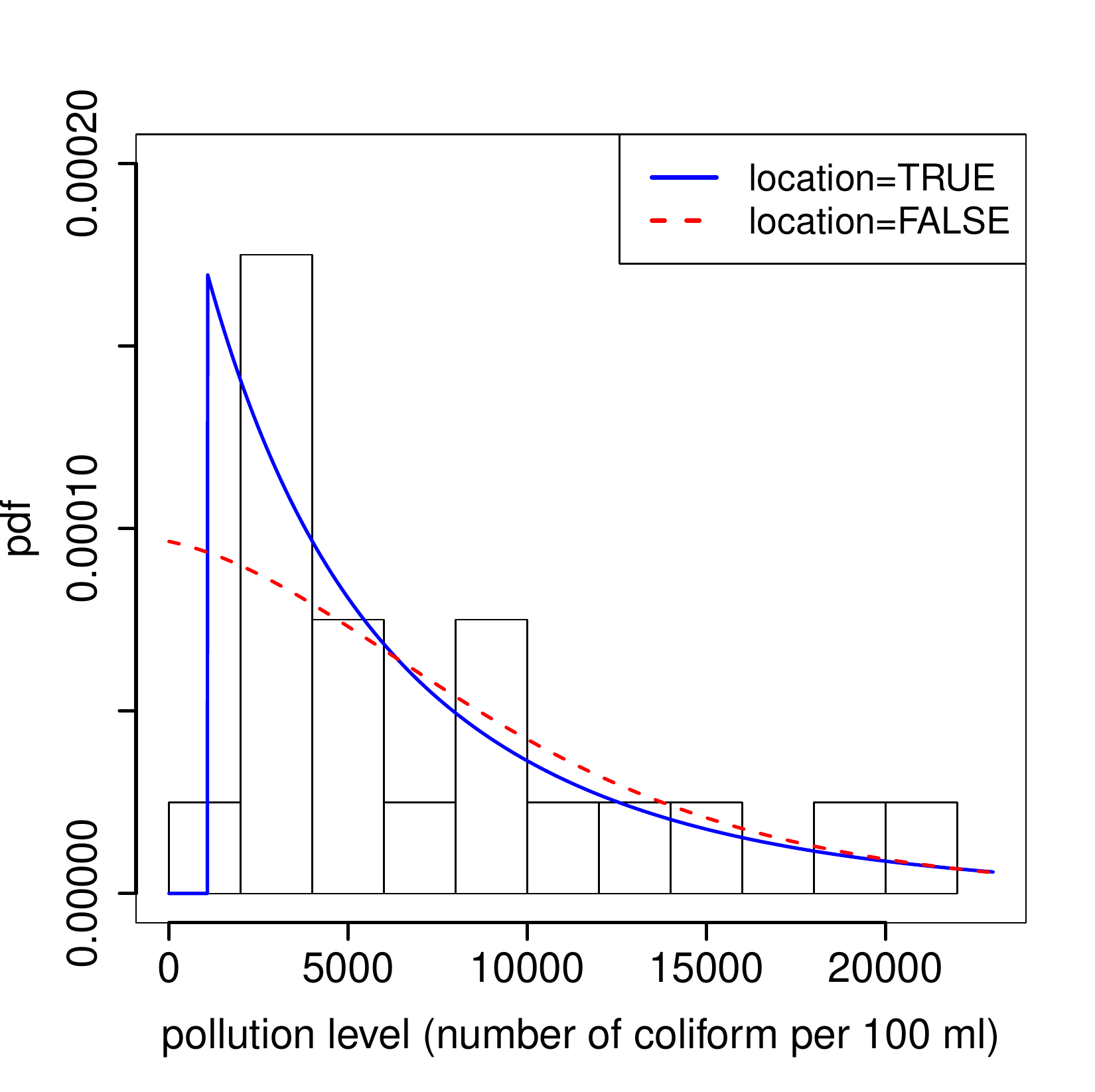}
\caption{pdf plot}
\end{subfigure}
\begin{subfigure}[b]{0.45\linewidth}
\includegraphics[width=\linewidth]{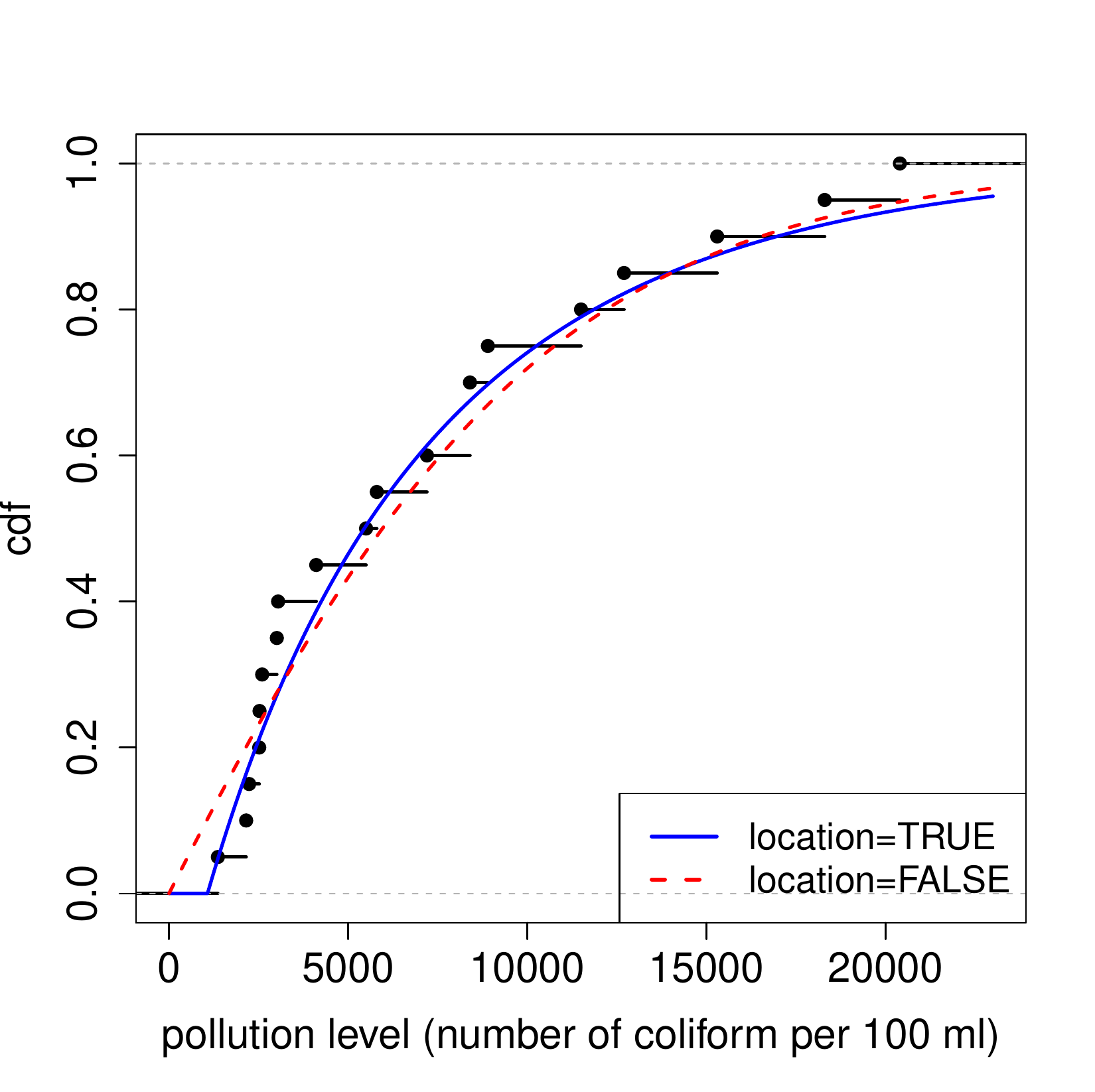}
\caption{cdf plot}
\end{subfigure}
\caption{Histogram and fitted probability density functions for pollution data are displayed in left hand-side subfigure. The empirical distribution function and fitted cumulative distribution function for pollution data are displayed is the right hand-side subfigure. In each subfigure two cases are considered. Those include the presence of the location parameter (location=TRUE) and absence of the location parameter (location=FALSE).}
\label{fig3}
\end{figure}

 Steen and Stickler (1976) reported the beach pollution level (measured in number of coliform per 100 ml) over 20 days in South Wales. As the third application, we apply the \verb|mog| family with cdf given in (\ref{mog}) to this set of data in two cases: 1- when $G$ follows a two-parameter exponential distribution with pdf given in (\ref{exppdf}), and 2- $G$ follows an exponential distribution. For this, we use the following commands.
\begin{verbatim}
R>x<-c(1364,2154,2236,2518,2527,2600,3009,3045,4109,5500,5800,7200,8400,8400,8900,
+ 11500,12700,15300,18300,20400)
R> mpsmog(x,"exp",TRUE,"Nedler-Mead",0.05)
$MPS
[1] 7.668608e-01 1.300979e-04 1.075007e+03

$Measures
      AIC     CAIC      BIC     HQIC         CM        AD       log    Moran
 395.7932 397.2932 398.7804 396.3763 0.04093273 0.2959096 -194.8966 78.72329

$KS
 statistic   p-value
 0.1241431 0.9175147

$`chi-square`
 statistic chi-value   p-value
  28.18183  31.41043 0.1051644

$`Convergence Status`
     [,1]                 
[1,] "Algorithm Converged"

R> mpsmog(x,"exp",FALSE,"Nedler-Mead",0.05)
$MPS
[1] 1.7785378951 0.0001715355

$Measures
      AIC     CAIC      BIC     HQIC         CM        AD       log    Moran
 398.5125 399.2183 400.5039 398.9012 0.06436965 0.4949034 -197.2562 80.00278

$KS
 statistic   p-value
 0.1508478 0.7529763

$`chi-square`
 statistic chi-value    p-value
  29.54727  31.41043 0.07752922

$`Convergence Status`
     [,1]                 
[1,] "Algorithm Converged"
\end{verbatim}
Clearly, when the location parameter is present, the \verb|mog| family yields a better fit for beach pollution data. Plots of histogram fitted density functions, empirical distribution function, and fitted distribution functions are displayed in Figure \ref{fig3}. This fact that presence of the location parameter yields a better fit for the beach pollution data is verified by statistics given in \verb|$Measure| part of related outputs.

\subsection{Random realization mechanism accuracy}
\begin{figure}[h!]
\centering
\begin{subfigure}[b]{0.45\linewidth}
\includegraphics[width=\linewidth]{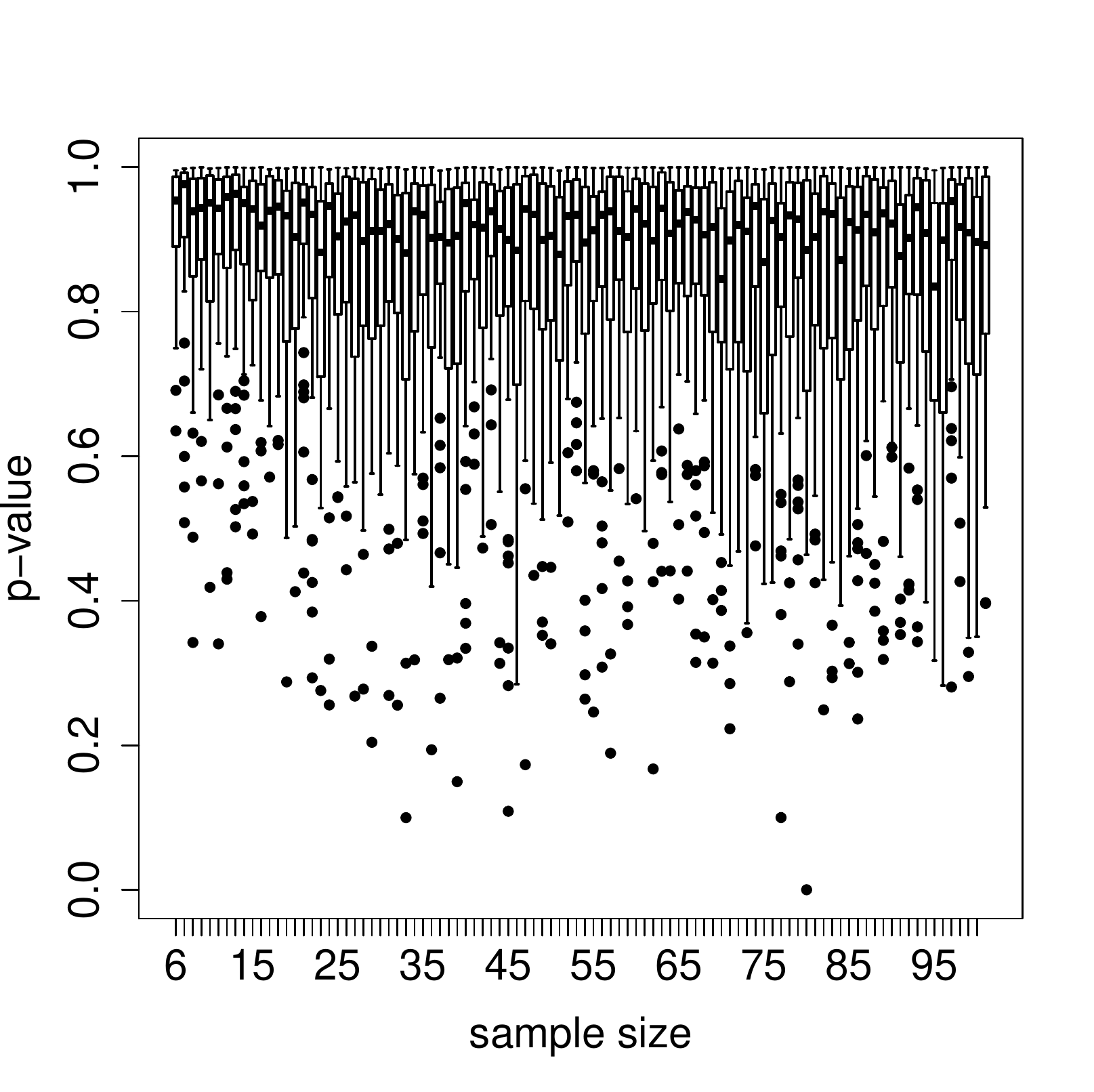}
\caption{Birnbaum-Saunders}
\end{subfigure}
\begin{subfigure}[b]{0.45\linewidth}
\includegraphics[width=\linewidth]{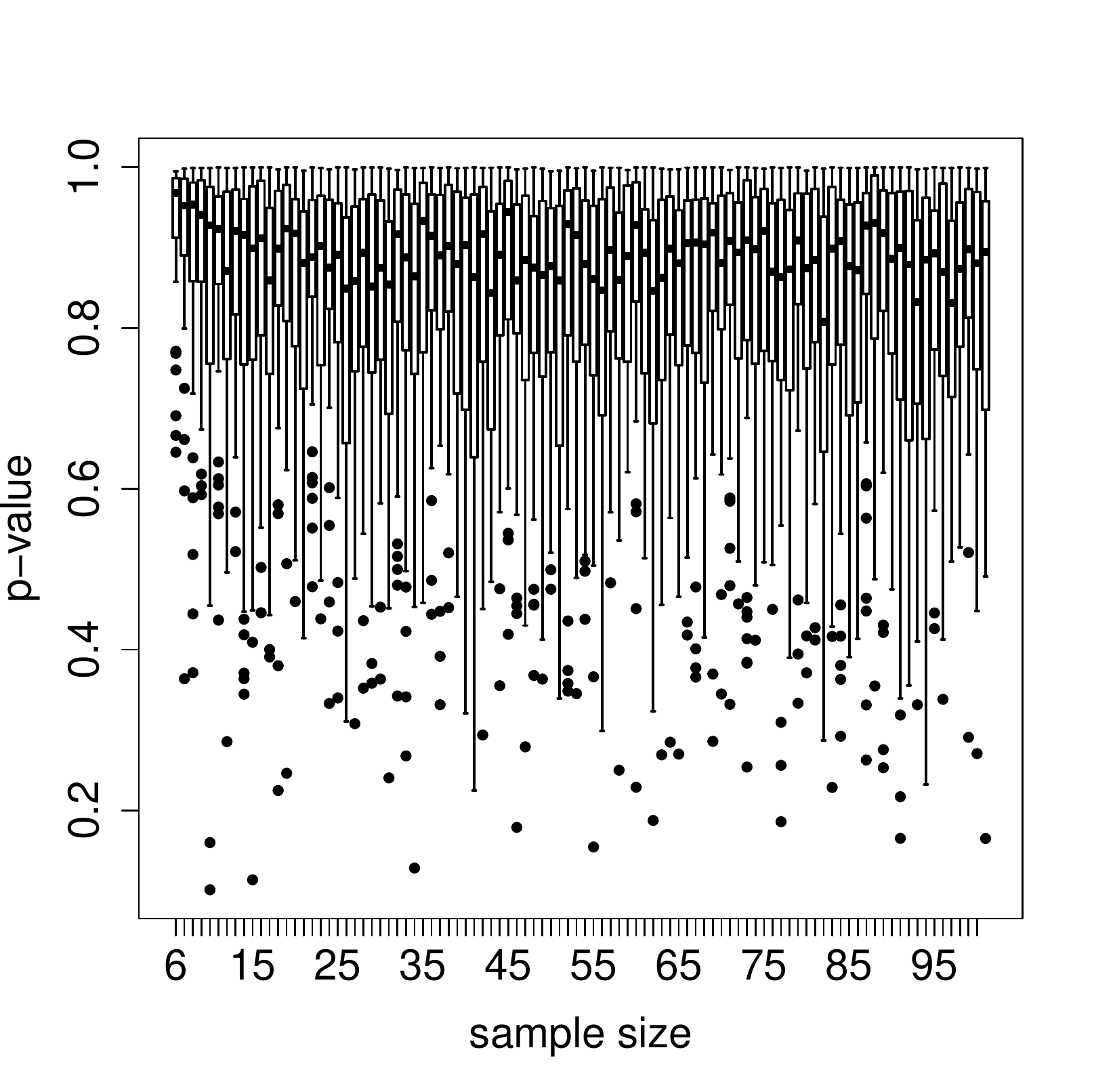}
\caption{Weibull}
\end{subfigure}
\begin{subfigure}[b]{0.45\linewidth}
\includegraphics[width=\linewidth]{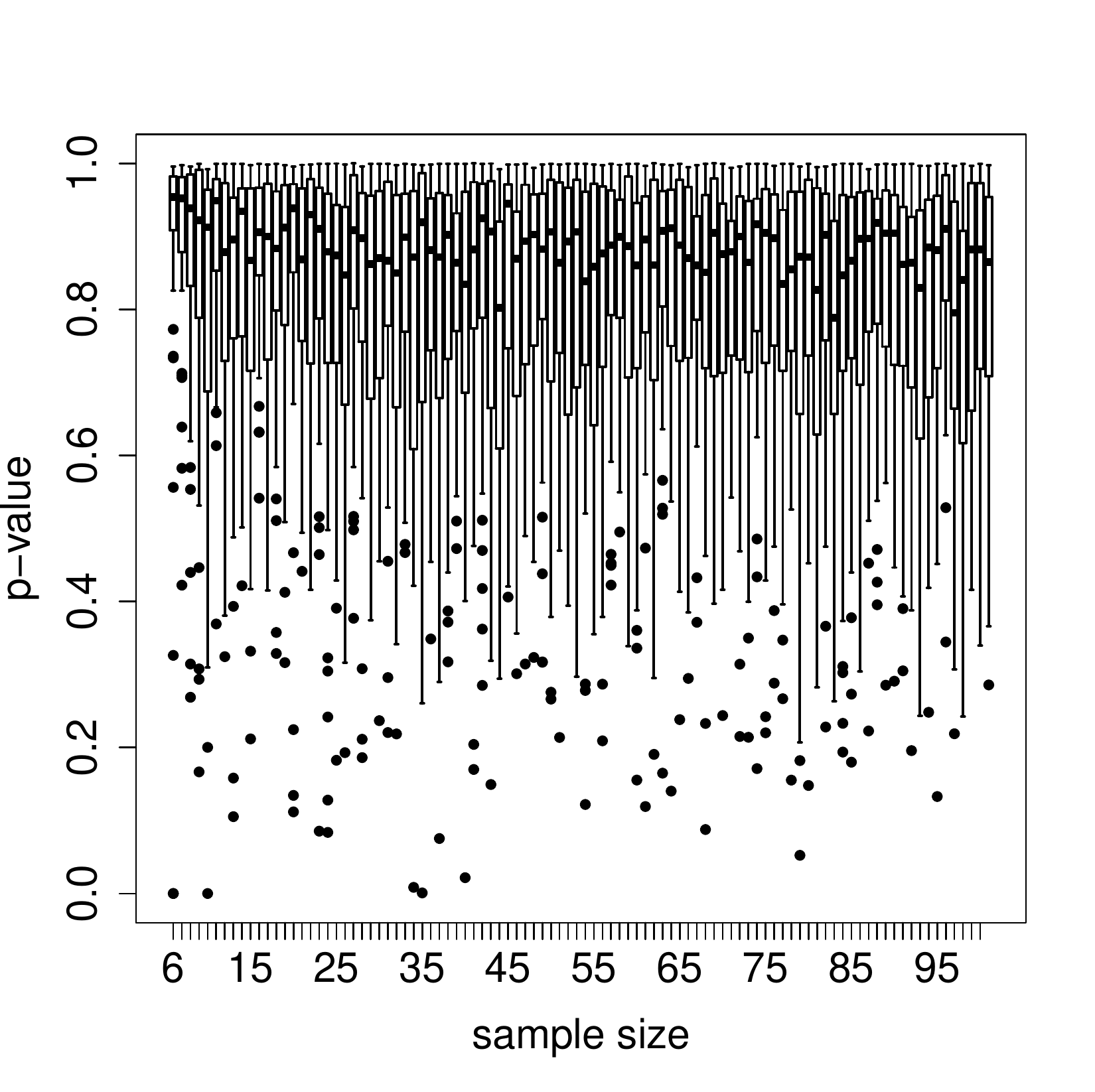}
\caption{lomax}
\end{subfigure}
\begin{subfigure}[b]{0.45\linewidth}
\includegraphics[width=\linewidth]{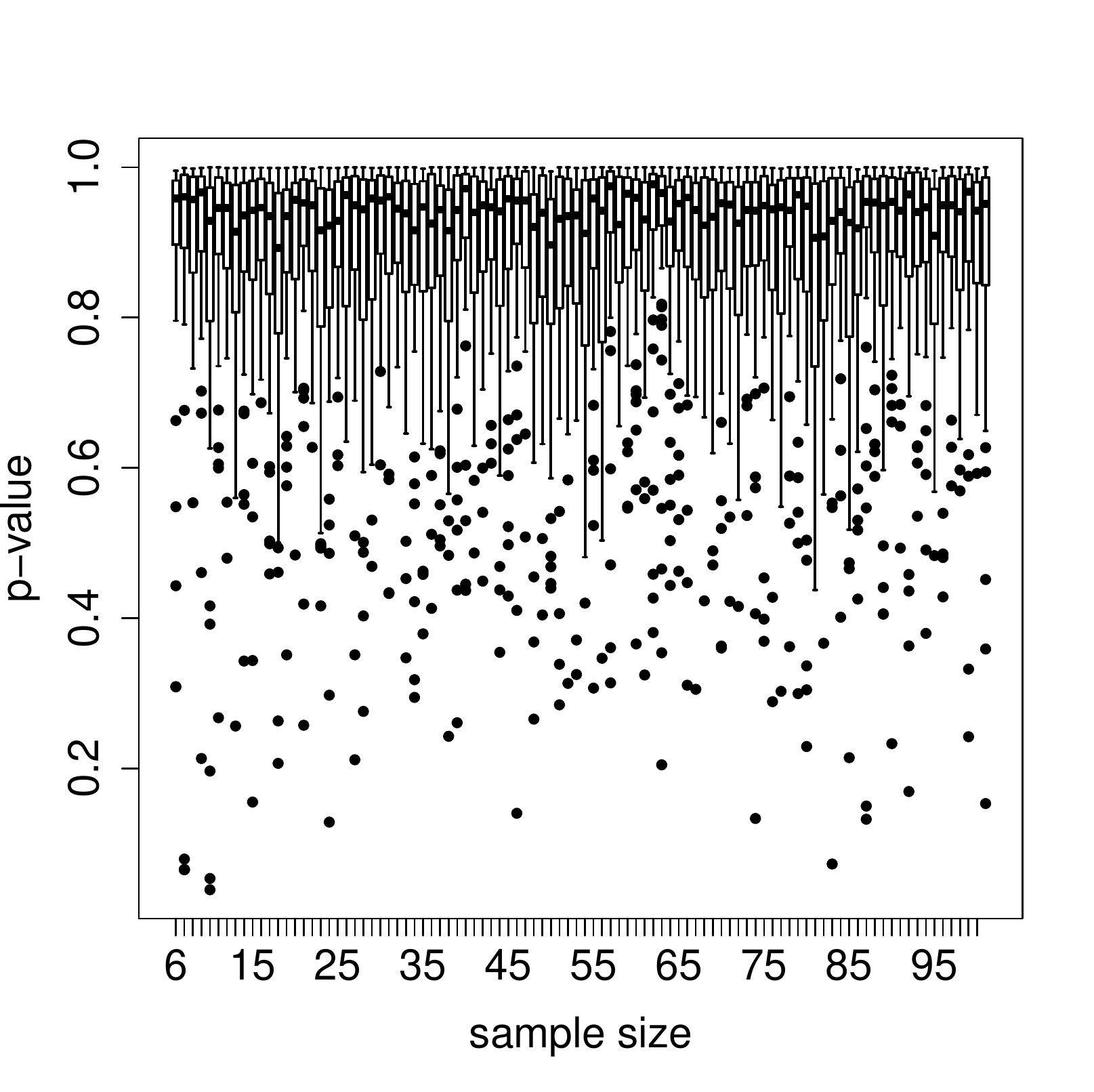}
\caption{log-logistic}
\end{subfigure}
\caption{Plots of p-value for testing the null hypothesis whether realizations come from the \texttt{loggammag1} family with specified $G$ or not. The name of $G$ distribution is presented under each subfigure.}
\label{fig4}
\end{figure}

Here, we perform a simulation study to check the accuracy of the \texttt{MPS} package for generating realizations from 24 $G$ families. To save the space, we confine ourselves to the study the simulation mechanism from the \verb|loggammag1| family when $G$ is one of \verb|"birnbaum-saunders"|, \verb|"log-logistic"|, \verb|"lomax"|, and \verb|"weibull"| distributions. For this aim, we follow the algorithm for each of four earlier mentioned $G$ distributions given by the following.
\begin{enumerate}
\item Generate a random sample of size $n$ from \verb|loggammag1| family using the routines provided in subsection \ref{simul},
\item Compute the p-value of the one-sample Kolmogorov-Smirnov hypothesis test that whether the sample follows the \verb|loggammag1| family distribution or not,
\item Repeat steps 1 and 2 for 100 times for each $n$ that ranges from 5 to 100, giving p-values $p_1, p_2,\dots, p_{100}$ say,
\end{enumerate}
The result of simulations are depicted in Figure \ref{fig4}. It should be noted that, for implementing the algorithm, all five components of $\widehat{\Theta}=(a, b,\alpha,\beta,\mu)^{T}$ are generated from uniform distribution over (0.5, 5) in each iteration. It follows, from Figure \ref{fig4}, that almost all of depicted boxplots are above 0.05 for all $n = 1, 2,\dots, 100$.
\section{Conclusion}
We have introduced an \verb|R| package, called \verb|MPS|, for statistical modelling of 24 shifted $G$ families of distributions. The statistical modelling involves computing the probability density function, computing the cumulative distribution function, computing the quantile function, simulating random realizations, and estimating the parameters via the maximum product spacings (MPS) approach introduced by Cheng and Amin (1983). The performance of \verb|MPS| package have been demonstrated through examples and real data applications. Adding a new shift (location) parameter to the 24 $G$ families of distributions made them more flexible and appropriate for modelling in practice. We have shown by the first real data application, when the maximum likelihood estimators break down, the MPS estimators exist and the \verb|MPS| package gives them. The \verb|MPS| package dose not depend on any other packages developed for \verb|R| environment and uploaded in Comprehensive R Archive Network (CRAN) at \url{https://cran.r-project.org/package=MPS}. The \verb|MPS| package can be updated for any new family of distributions in the future.

\end{document}